\documentstyle[amssymb]{article}
\def\Im{\mathop{\mbox{\rm Im}}\nolimits}
\def\pa{\partial}
\def\La{\Lambda}
\def\h{\hbar}
\def\va{\varphi}

\def\La{\Lambda}
\def\CP{{\cal P}}
\def\CK{{\cal K}}
\def\CH{{\cal H}}
\def\CF{{\cal F}}
\def\CA{{\cal A}}
\def\ve{\varepsilon}
\def\ga{1+\gamma^{-1}}
\def\sp{\langle\vec\sigma,\vec\CP\rangle}
\def\sb{\langle\vec\sigma,\vec\beta\rangle}
\def\bp{\langle\vec\beta,\vec\CP\rangle}
\def\sbb{\langle\vec\sigma,\vec\beta\times\dot{\vec\beta}\rangle}
\def\sdb{\langle\vec\sigma,\dot{\vec\beta}\rangle}
\def\bdb{\langle\vec\beta,\dot{\vec\beta}\rangle}
\def\bve{\langle\vec\beta,\vec E\rangle}
\def\bvet{\langle\vec\beta,\vec E(t)\rangle}
\def\sve{\langle\vec\sigma,\vec E\rangle}
\def\svet{\langle\vec\sigma,\vec E(t)\rangle}
\def\bvh{\langle\vec\beta,\vec H\rangle}
\def\bvht{\langle\vec\beta,\vec H(t)\rangle}
\def\svh{\langle\vec\sigma,\vec H\rangle}
\def\svht{\langle\vec\sigma,\vec H(t)\rangle}
\def\sbe{\langle\vec\sigma,\vec\beta\times\vec E\rangle}
\def\sbet{\langle\vec\sigma,\vec\beta\times\vec E(t)\rangle}

\def\rot{\mathop{\mbox{\rm rot}}\nolimits}
\def\div{\mathop{\mbox{\rm div}}\nolimits}
\def\grad{\mathop{\mbox{\rm grad}}\nolimits}
\def\eg{\frac{e(g-2)\h}{4m_0c}}
\def\lan{\langle}
\def\ran{\rangle}

\makeatletter
\@addtoreset{equation}{section}
\makeatother

\newtheorem{Prop}{Property}%
\newtheorem{Def}{Definition}%
\newtheorem{theo}{Theorem}%
\newtheorem{Lem}{Lemma}%
\newtheorem{rem}{Remark}%

\begin{document}

\title{Semiclassical trajectory-coherent approximation in quantum
mechanics: II. High order corrections to the Dirac operators in
external electromagnetic field}
\author{V.G.Bagrov$^1$ \thanks {e-mail: bagrov@hcei.tomsk.su},
V.V.Belov$^2$ \thanks {e-mail: belov@amath.msk.ru},
and A.Yu.Trifonov$^3$ \thanks {e-mail: trifonov@phtd.tpu.edu.ru}}
\date{$^1${\it High-Current Electronics Institute,\\
Siberian Division Russian Academy of Science\\
4 Akademichesky Ave., 634055 Tomsk Russia\\
$^2$Department of Applied Mathematics\\
Moscow Institute of Electronic and Mathematics,\\
B. Vusovsky 3/12, 109028 Moscow, Russia\\
$^3$Department of Mathematical Physics,\\
Tomsk Polytechnical University, 30 Lenin Ave., 634034 Tomsk, Russia}}
\maketitle

\begin{abstract}
High approximations of semiclassical trajectory-coherent states (TCS) and
of semiclassical Green function (in the class of semiclassically
concentrated states) for the Dirac operator
with anomalous Pauli interaction are obtained. For
Schr\"odinger and Dirac operators  trajectory-coherent
representations are constructed up to any precision with respect
to $\hbar$, $\hbar\to0$.
\end{abstract}

\section*{Introduction}

This paper is the second part of our work \cite{1} and
deals with  higher approximations of semiclassical
trajectory-coherent states (TCS) for the Dirac equation with
anomalous Pauli interaction \cite{2} in an external
electromagnetic field. The approach given in \cite{3} is based
on the concept that classical equations of motion can be
considered as the limiting (as $\hbar\to0$) equations of motion
for averaged values of corresponding quantum mechanical magnitudes.
In physical literature this passage to the limit is justified by
describing the dynamical states of a quantum system as
wave packets localized near the position of a classical
particle. For more detailed references, see paper \cite{1} and
reviews \cite{5, 6}. Such states can be obtained up to
any power of $\hbar^{1/2}$, $\hbar\to0$, by the complex
WKB-Maslov method called the complex germ theory \cite{4}.

In [1, 7--9], complete orthonormalized sets of asymptotic (as
$\hbar\to0$) semiclassical solutions to wave equations
(Schr\"odinger, Klein--Gordon, pseudodifferential equation of
the Schr\"odinger type for a relativistic particle) were constructed
for a scalar quantum particle in an arbitrary external field.
These sets satisfy the coherence condition: {\it the quantum mechanical
values\/} ({\it averaged with respect to the functions from these sets\/})
{\it for the operators of coordinates and momenta are\/} ({\it as
$\hbar\to0$\/}) {\it solutions of classical equations of motion\/}
({\it Newton and Lorentz respectively\/}).

Such solutions, called semiclassical trajectory-coherent states
(TCS), generalize the well-known coherent [10--14] and
correlated  coherent states of nonrelativistic quadratic systems
\cite{15} to the case of an arbitrary external field and scalar
relativistic equations.

For a relativistic quantum particle in an arbitrary external
field, which is described by the Dirac equation, the
semiclassical TCS were constructed in [16--18]. They are
characterized by the fact that, beside the coherence condition
with respect to orbital variables, they must also satisfy the
following condition of spin coherence:
{\it in the limit as $\hbar\to0$, the quantum mechanical averages of
spin operator, i.e., of the Bargman polarization vector $\hat
S^\mu$ {\rm \cite{19}}, are solutions to the classical
relativistic equations of spin motion, which are the
Bargmann--Michel--Telegdi equations\/} \cite{20}.

Note that the construction of semiclassical TC-states for
quantum systems is based on the semiclassical trajectory
coherent representation of the corresponding Hamiltonians. For
scalar quantum mechanical equations, this representation was
implicitly used in essence in \cite{8}. For the Dirac operator,
it was introduced first in \cite{16, 17}. From the viewpoint of
semiclassical asymptotics, the constructed TC-states give the
principal term of asymptotic expansion in the powers of
$\h^{1/2}$, $\h\to0$. They satisfy the corresponding
evolutionary quantum mechanical equations up to the functions,
whose $L_2$-norm is of the order $O(\h^{3/2})$ as $\h\to0$
($\bmod O(\h^{3/2})$). In this paper we construct the higher
approximations of the trajectory-coherent states of the Dirac
operator in an arbitrary magnetic field, i.e., the asymptotic
solutions of these equations up to $O(\h^{(N+1)/2})$, where
$N$ is arbitrary integer independent of $\h$; $N=3,4,\dots$.

In the general situation the scheme for construction of
corrections to the principal term of a semiclassical asymptotic
solution is well know \cite{4}. Nevertheless, its realization
for particular quantum equations with arbitrary electromagnetic
potentials requires a large amount of nontrivial calculations.
In particular, we succeeded in constructing a unitary (up to any
order in $\h^{1/2}$) operator in the form of an asymptotic
series in powers of $\h^{1/2}$, $\h\to0$, which
defines the passage to the semiclassical trajectory-coherent
representation.

In the semiclassical description of the behavior  of a quantum
particle with its spin properties taken into account, this
allowed to construct the two-component theory, i.e.,
to use the space of positive-frequency solutions to the Dirac
equations, to exclude explicitly  the Hamiltonian which is a
relativistic generalization of the Pauli operator \cite{21}.
The scalar part of this Hamiltonian describes quantum fluctuations
of the wave packet near the particle position on a classical
trajectory, and its ``vector'' part describes the interaction
between the particle spin and the external field and the quantum
fluctuations with respect to orbital variables.

Note that in order to study certain specific physical problems,
e.g., to consider with a given precision as $\h\to0$ \cite{22, 23}
the quantum effects which appear when a charge radiates spontaneously
in an arbitrary external field, we must take into account the higher
(as $\h\to0$) corrections to the principal term of TCS. In this
case, to obtain the first (up to $O(\h^{2})$) quantum corrections to
the power of radiation (which are uniform with respect to relativism, i.e.,
they can be used in any region of the particle energy), it is necessary
to take the semiclassical TC-states satisfying the Dirac equation
up to $O(\h^{5/2})$.

The influence of corrections to the principal term of asymptotics
is essential when  nonadiabatic phases in quantum mechanics are
considered \cite{24}, as well as when equations of motion of
quantum averages are derived [1, 25--27].

This paper is organized as follows. Section 1 contains some
relevant facts about the semiclassical trajectory-coherent
representation (TC-representation) for the Schr\"odinger
equation. In Section 2 we give the definition and investigate the simple
properties of the semiclassically concentrated states of the Dirac
equation. Higher approximations of semiclassical trajectory-coherent
states $\Psi_\nu^{(N)}(\vec x,t,\h)$ and of semiclassical Green function
$G_D^{(N)}(\vec x,\vec y,t,s)$ (in the class of semiclassically concentrated
states) for the Dirac equation are constructed in Sections 3 and 4. In
Section 5 we show that in the class of positive-frequency (negative-frequency)
$\sqrt{\h}$, $\h\to0$, semiclassically concentrated states it can be gone
to the one particle two-component theory whose Hamiltonian
$\widehat\CH_D^{(N)}(t)$ is a self-ajoint operator for any order of
$\h\to0$. Portion of necessary material is taken into appendices.
Our paper is essentially based on the results and notation of Part 1
\cite{1}, some of them are not defined in this text, but only referred to as, e.g., (I.2.3),
which means formula (2.3) in the first part \cite{1}.

\section{Semiclassical trajectory-coherent representation \protect\newline
for the Schr\"odinger equation}

As is known, physical results in quantum theory are independent
of the choice of the representation for its principal dynamical
variables. A good choice of one or another representation often
allows to simplify the problem or to solve it completely. To
solve the problem of passing in the limit from quantum
to classical mechanics,  a new semiclassical trajectory-coherent
(TC) representation was constructed in \cite{16}  up to
$\bmod \,\hat O(\h^{1/2}) $, $\h\to 0$. In this
representation, in the limit as $\h\to 0$, the quantum
averages for an arbitrary observable $\hat A_\h(t)$ with
classical analog $\hat A_\hbar (t)= A(\hat{\vec p},\hat{\vec x},t)$
go over into the classical observable $A(\vec p,\vec x,t)$
translated along the trajectories of the corresponding Hamilton
system (I.1.1). In concrete problems, from the viewpoint of
applications, it is important to know how the quantum-mechanical
averages tend to their classical limit, i.e., the corresponding
asymptotic expansions must be obtained up to any precision with
respect to $\h$, $\h\to0$. To solve this problem, we shall
construct a semiclassical trajectory-coherent representation for
the Schr\"odinger equation (I.0.1) up to $\hat O(\h^{(N+1)/2})$,
$N=1,2,\dots$. In Section 4, in order to construct a semiclassical
TC-representation of the Dirac operator in an arbitrary external
field, we shall use the construction of a unitary operator (with
a given precision in $\h\to0$), which defines the passage to
such representations.

Let us define the Hilbert space of functions depending on $\vec
x$ and $\h$ with the following scalar product of
two functions $\va_1 (\vec x,t,\h)$ and $\va_2 (\vec x,t,\h)$:
\begin{equation}
\langle {\va_1}|{\va_2}\rangle_{L_\h^t}=\int\limits_{{\Bbb R}^3}\va_1^*
(\vec x,t,\h)\va_2(\vec x,t,\h)\rho_\h^{z_0}(\vec x,t)\,d^3 x,
\end{equation}
where the normalized measure density $\rho_\h^{z_0}(\vec x,t)$ is equal to
\begin{equation}
\rho_\h^{z_0}(\vec x,t)=N_\h^2|J(t,z_0)|^{-1}\exp\Big\{
\frac{2}{\h} \Im\,S(\vec x,t)\Big\}.
\end{equation}
Here the complex action $S(\vec x,t)$ and the normalizing factor
$N_{\h}$ are defined in (I.1.11). It is natural to consider the
space $L_\h^t$  as the space of states of quantum system (I.0.2) which,
as $\h\to0$, are localized in a neighborhood of the position
$z(t, z_0)$ of a classical particle on the phase trajectory,
since $\rho_\h^{z_0}(\vec x,t)\to \delta(\vec x-\vec x(t))$ as
$\h\to0$ and, in the $p$-representation, we have
$\hat F^\h_{x\to p}\rho_\h^{z_0}(\vec x,t)\to \delta(\vec p-\vec p(t))$,
where $\hat F^\h_{x\to p}$ is the $\h^{-1}$-Fourier
transform \cite{28}. The measure depends on a small parameter
which implies that one can approximate (up to a given precision
$\h\to0$) the smooth functions $\varphi(\vec x, t)$ by partial
sums of the Taylor series in powers of $\Delta \vec x$:
$$
\va(\vec x,t)=\sum_{k=0}^{N}\frac{1}{k!}d^k\va(t)+R_N(\vec x,t),
$$
where $d^k\va(t)$ is the $k$-th term of the expansion of
$\va(\vec x(t,z_0)+\Delta\vec x,t)$ into a Taylor
series in powers of $\Delta \vec x$, and for $R_N(\vec x,t)$ we have
the estimate: $\|R_N(\vec x,t)\|_{L_\h^t}=O(\h^{(N+1)/2})$.
This estimate follows from the relation
$\int\rho_\h^{z_0}(\vec x,t)\langle \vec a,\Delta \vec x\rangle^{N+1}
\,d\vec x=O(\h^{(N+1)/2})$ which holds for any constant vector $\vec a$
(comp. with (I.4.8)). On the other hand, the space $L_\h^t$
contains, as its elements, the functions $\va(\vec x,t,\h)$ which
depend on $\h\to0$ in a singular way and are of the order
$O(1)$ as $\h\to0$ in the norm of $L^t_\h$. For example,
$$
\va(\vec x ,t,\h)=\Big(\frac{\langle\vec a(t),\Delta \vec x\rangle}
{\sqrt{\h}}\Big)^k.
$$
Denote by $\CP_\h$ the set of polinomials in powers of $\h^{-1/2}\Delta\vec x$
with coefficientss depending on $t$. The set $\CP_\h$ is everywhere dense
in $L^t_\h$ and is natural domain of definition for differential operators
acting in $L^t_\h$. By $\hat O(\h^\alpha)$ we shall denote the operator
$\hat F(\h):L_\h^t\to L_\h^t$ for which on the set $\CP_\h$ the estimate holds:
$\|\hat F(\h)\va\|_{L_\h^t}=O(\h^\alpha)$ as $\h\to0$ uniformly in $t=[0,T]$.
Note that in this sense the operators
\begin{equation}
\sqrt{\h}\nabla,\quad\langle \Delta\vec{x}, \nabla \rangle,\quad
\frac{d}{dt}=\frac{\pa}{\pa t}+\langle\dot{\vec x} (t,z_0),
\nabla\rangle
\end{equation}
are the order $\hat O(1)$ as $\h\to0$.

Let us define the operator $\hat\CK_S^{(N)}(t,\h):L_\h^t\to L_2({\Bbb R}_3^n)$,
which defines (up to $\bmod\hat O(\h^{(N+1)/2}))$ the passage to the semiclassical
TC-representation by the formula:
\begin{eqnarray}
\hat\CK_S^{(N)}(t,\h)\va&=&\sum_{n=0}^N\Big[-\frac{i}{\h}
\hat\CK_1(t)\Big]^n\hat\CK_S^{(0)}(t,\h)\va, \cr
\hat\CK_S^{(0)}(t,\h)&=&\frac{N_\h}{\sqrt{J(t,z_0)}}
\exp\Big[\frac{i}{\h}S(\vec x,t)\Big],\\
\hat\CK_1(t)\va(t)&=&\sum_{|\nu|=0}^\infty|\nu,t\rangle
\int\limits_0^t d\tau\langle\tau,\nu|\hat\CH_1(\tau)|\va(\tau)\rangle,\nonumber
\end{eqnarray}
where the functions $|\nu,t\rangle$, $S(\vec x,t)$, $J(\vec x,t)$ and the constant
$N_\h$ are defined in (I.1.11)--(I.1.15), and the operator $\hat\CH_1$ is
defined in (I.4.7).

The operator $\hat\CK_S^{(N)}(t,\h)$ maps the space $L^t_\h$ into $L_2$
unitarily up to $O(\h^{(N+1)/2})$ which means that
\begin{equation}
\langle\hat\CK_S^{(N)}(t,\h)\va_1|\hat\CK_S^{(N)}(t,\h)\va_2\rangle_{L_2}=
\langle\va_1|\va_2\rangle_{L_\h^t}+O(\h^{(N+1)/2}).
\end{equation}

The Schr\"odinger equation (I.0.1) in the semiclassical TC-representation
given the operator $\hat\CK_S^{(N)}(t,\h)$ (1.4) has the form
\begin{eqnarray}
&\big[\hat\CK_S^{(N)}(t,\h)\big]^{-1}\big\{-i\h\pa_t +\hat\CH\big\}
\hat\CK_S^{(N)}(t,\h)\va=\hat\pi_0\va +O(\h^{(N+3)/2}),\\
&\hat\pi_0=(-i\h)\big\{\displaystyle\frac{d}{dt} + \langle\Delta\vec x,
[\CH_{xp}(t)+Q(t)\CH_{pp}(t)]\nabla\rangle
-\frac{-i\h}{2}\langle\nabla,\CH_{pp}(t)\nabla\rangle\big\},\\
&\displaystyle\frac{d}{ dt}=\frac\pa{\pa t}+\langle\dot{\vec x}(t,z_0),\nabla\rangle,\nonumber
\end{eqnarray}
and $Q(t)$ is defined in (I.1.9). Thus with precision up to functions of
the order $O(\h^{(N+3)/2})$, $N=0,1,2,\dots$, in the norm of the space
$L^t_\h$, the Schr\"odinger equation in the semiclassical TC-representation
is equivalent to the equation
\begin{equation}
\hat\pi_0\va =0, \quad \va\in L_\h^t.
\end{equation}
Equation (1.8) can be easily integrated if we note that the operator $\hat\pi_0$
admits the complete set of dynamical symmetries. It is easy to verify that the
operators
\begin{equation}
\begin{array}{l}
\hat\La_j^+=\displaystyle\frac{1}{\sqrt{2\h\,\Im\,b_j}}\big(
\langle\vec Z_j^*(t),\hat{\vec p}\rangle-\langle(\vec W_j^*-Q(t)
\vec Z_j^*(t)),\Delta\vec x\rangle\big),\\[6pt]
\hat\La_j=\displaystyle\frac 1 {\sqrt{2\h\,\Im\,b_j}}\langle\vec Z_j(t),
\hat{\vec p}\rangle, \quad j=\overline{1,n},\end{array}
\end{equation}
commute with the operator $\hat\pi_0$ and satisfy the Bose commutation relations \cite{4}:
$$
\big[\hat\La_k,\hat\La_j^+\big]=\delta_{kj},\quad
\big[\hat\La_k,\hat\La_j\big]=\big[\hat\La_k^+,\hat\La_j^+\big]=0.
$$
Hence the functions $H_\nu=|H_\nu\rangle=\prod\limits_{j=1}^n(\nu_j!)^{-1/2}
(\hat\La_j^+)^{\nu_j} 1$ form in $L_\h^t$ the complete orthonormalized set of
solutions of equation (1.8). Applying the operator $\hat\CK_S^{(N)}(t,\h)$
to the functions $|H_\nu\rangle$, we obtain the higher approximations for
semiclassical TCS (I.4.14) of the Schr\"odinger equation.

If follows from (1.6), (1.9) thaat the operators
\begin{eqnarray*}
&\hat\La_j^{(N)}=\hat\CK_S^{(N)}(t,\h)\hat\La_j
\big[\hat\CK_S^{(N)}(t,\h)\big] ^{-1},\\
&\big[\hat\La_j^+\big]^{(N)}=\hat\CK_S^{(N)}(t,\h)\hat\La_j^+
\big[\hat\CK_S^{(N)}(t,\h)\big]^{-1}, \quad j=\overline{1,n}
\end{eqnarray*}
are an approximate symmetry operators for Schr\"odinger equation
$$
\Big[\big\{-i\h\pa_t+\hat{\CH}\big\},
\hat\La_j^{(N)}\Big]=\Big[\big\{-i\h\pa_t+\hat{\CH}\big\},
\big[\hat\La_j^+\big]^{(N)}\Big]=\hat O(\h^{(N+3)/2})
$$
and satisfy the following commutation relations:
\begin{eqnarray*}
\Big[\big[\hat\La_j^+\big]^{(N)},\big[\hat\La_k^+\big]^{(N)}\Big]
=\Big[\hat\La_j^{(N)},\hat\La _k^{(N)}\Big]=\hat O(\h^{(N+3)/2}), \\
\Big[\hat\La_j^{(N)},\big[\hat\La_k^+\big]^{(N)}\Big]=\delta_{jk}+
\hat O(\h^{(N+3)/2}).
\end{eqnarray*}

For $N=2$ we calculate the operator $\hat\CK_S^{(N)}(t,\h)$ explicitly.
By (1.4) and the nonrelativistic Hamiltonian function
$$
\CH= \frac 1{2m}\vec\CP^2 +e\Phi, \quad
\vec\CP=\vec p-\frac e c \vec\CA,
$$
we have
\begin{equation}
\hat\CK_S^{(2)}(t,\h)=\hat\CK_S^{(0)}(t,\h)\big[1-i\sqrt{\h}
\hat\pi_1-i\h\hat\pi_2-\h\hat\pi_1^2\big],
\end{equation}
where
\begin{eqnarray}
\h^{j/2}\hat\pi_j\varphi(t)&=&\frac{1}{(j+2)!}
\sum_{|\nu|=0}^{\infty} | H_{\nu}\rangle
\int\limits_0^t d\tau\langle H_{\nu}|\frac{1}{\h}\hat D^{j+2}
\CH(\tau)|\varphi(\tau)\rangle;\quad j=1,2,\cr
\frac{1}{3!}\hat D^3\CH(t)&=&-\frac{e}{4mc}(\langle\hat{\vec
\CP}{}_1',d^2{\vec\CA}\rangle+\langle d^2{\vec\CA},
\hat{\vec\CP}{}_1'\rangle)+\frac{e}{3!}(d^3{\Phi}-\langle
\vec\beta,d^3\vec\CA\rangle),\cr
\frac{1}{4!}\hat D^4\CH(t)&=&\frac{e}{4!}( d^4\Phi - \langle
\vec{\beta}, d^4{\vec\CA}\rangle) + \frac{e^2}{8mc^2}\langle
d^2{\vec\CA},d^2{\vec\CA}\rangle-{}\\
&&{}- \frac{e}{2mc}\frac{1}{3!}(\langle\hat{\vec\CP}{}_1',
 d^3{\vec\CA}\rangle+\langle d^3{\vec\CA},\hat{\vec\CP}{}'_1\rangle),\cr
\hat{\vec\CP}{}'_1&=&-i\h\nabla+Q(t)\Delta\vec{x} -
\frac{e}{c} d{\vec\CA (t)},\quad \vec \beta = \frac 1 c \dot{\vec x}
(t,z_0),\nonumber
\end{eqnarray}
and $d^k\CA(t)$ denotes the function $d^k\CA(t)=(\langle
\Delta\vec x,\pa/\pa y\rangle )^kA(\vec y,t)|_{\vec y=\vec x(t,z_0)}$.
Note that since the operators $\hat{\vec\CP}{}'_1$ and $\Delta\vec x$
are slef-adjoint in $L_\h^t$ formulas (1.11) define the self-adjoint in
$L_\h^t$ operators $\hat D^j \CH (t)$, $j=3,4$. In particular, this implies
that the operator $\hat\CK_S^{(N)}(t,\h)$ is unitary for $N=2$ (up to
$\hat O(\h^{3/2})$):
\begin{eqnarray}
\lefteqn{\big\langle\big(\hat\CK_S^{(2)}(t,\h)H_{\nu'}\big)|
\big(\hat\CK_S^{(2)}(t,\h)H_\nu\big)\big\rangle_{L_2} =
\big\langle H_{\nu'}|\big(\hat\CK_S^{(2)}(t,\h)\big)^+
\hat\CK_S^{(2)}(t,\h)|H_\nu\big\rangle_{L_\h^t}=}\cr
&&=\langle H_{\nu'}|\big[1-i\sqrt{\h}(\hat\pi_1-\hat\pi_1^+)- i\h
(\hat\pi_2 - \hat\pi_2^+) +\h(\hat\pi_1^+ \hat\pi_1-\hat\pi_1^2 -(\hat\pi_1^2)^+)
\big]|H_\nu\rangle+O(\h^{3/2})= \cr
&&\qquad\qquad\qquad\qquad\qquad=\delta_{\nu,\nu'} + O(\h^{3/2}).
\end{eqnarray}

By using (1.10) for any arbitrary operator $\hat A_\h (t)=A(\hat{\vec p},
\hat{\vec x},t)$ with Weyl symbol $A(\vec p,\vec x,t)$ possessing a classical
analog, we can find its explicit form (up to $\hat O(\h^{3/2})$) in
the TC-representation
\begin{eqnarray}
\lefteqn{\hat A_\h'=(\hat\CK_S^{(2)}(t,\h))^{-1}
\hat A_\h(t)\hat\CK_S^{(2)}(t,\h) +\hat O(\h^{3/2})={}}\cr
&&=A(t)+\hat D^1A(t) + \frac{1}{2}\hat D^2A(t)-i\sqrt{\h}
\big[(\hat D^1A(t))\hat\pi_1 - \hat\pi_1^+\hat D^1A(t)\big] +
\hat O(\h^{3/2}),
\end{eqnarray}
where by $\hat D^j A(t)$, similarly (I.4.8), we denote the operator of the form
\begin{equation}
\begin{array}{l}
\hat D^jA(t) = {\Big(\langle\Delta\hat{\vec p}\,{}',\displaystyle\frac{\pa}
{\pa\vec z}\rangle + \langle\vec{\Delta x},\displaystyle\frac{\pa}
{\pa\vec y}\rangle\Big)}^jA(\vec z,\vec y,t)\big|_{\vec z =
\vec p (t,z_0),\vec y = \vec x (t,z_0)},\\
\Delta\hat{\vec p}{}'=- i\h\nabla + Q(t)\Delta\vec{x},\end{array}
\end{equation}
and the function $A(t)$ is the classical observable $A(\vec p,\vec x,t)$,
corresponding to the operator $\hat A_\h (t)$, calculated in the point
$z(t,z_0)$ on a phase trajectory, $A(t)=A(\vec p(t,z_0),\vec x(t,z_0),t)$.

\section{Theorem about semiclassicaly concentrated states of the
Dirac equation}

Let us consider a relativistic particle described by the Dirac
equation  with  anomalous Pauli interaction
\begin{equation}
\hat L_D\Psi = 0, \qquad \hat L_D=-i\h\pa_t+\hat\CH_D,
\end{equation}
where Hamiltonan $\hat\CH_D$ has the form \cite{29}:
\begin{equation}
\begin{array}{l}
\hat\CH_D = \hat\CH_0 + (-i\h)\hat\CH_1,  \quad
\hat\CH_0 = c\langle\vec\alpha,\hat{\vec\CP}\rangle +
\rho_3m_0c^2 + e\Phi(\vec x,t),\\[6pt]
\hat\CH_1=\displaystyle\frac{ ie_0(g-2)}{4m_0c}\big[\rho_3\langle
\vec\Sigma,\vec H(\vec x,t)\rangle+\rho_2\langle\vec\Sigma,
\vec E(\vec x,t)\rangle\big],\\[6pt]
\vec\alpha = \rho_1\vec\Sigma,\quad\vec\Sigma=\left(\begin{array}{cc}
\vec\sigma & 0 \\0 & \vec\sigma\end{array}\right),\quad e_0 = -e,\\
\rho_1 =\left(\begin{array}{cc}0&{\bold I}\\{\bold I}&0\end{array}\right),\quad
\rho_2 = \left(\begin{array}{cc} 0&- i{\bold I}\\ i{\bold I}&0\end{array}
\right),\quad\rho_3 = \left(\begin{array}{cc}{\bold I}&0\\0&-{\bold I}\end{array}
\right).\end{array}
\end{equation}
Here $\vec\sigma=(\sigma_1,\sigma_2,\sigma_3)$ are the Pauli matrices, 0 and
{\bf I} are the zero and iunit $2\times 2$ matrices,
$\vec H(\vec x,t)=\rot\,\vec\CA(\vec x,t)$ is a magnetic field,
$\vec E(\vec x,t) = -\nabla\Phi(\vec x,t)-c^{-1}\pa_t\vec\CA(\vec x,t)$
is the electric field, $g$ is the gyromagnetic ratio.

Precisely as in the nonrelativistic case, we assume that the
electromagnetic potentials $\Phi(\vec x,t)$ and $\vec A(\vec x,t)$ are
smooth functions of $\vec x \in{\Bbb R}^3$, $t\in {\Bbb R}^1$ and as
$|\vec x|\to\infty$ increase uniformly in $t\in{\Bbb R}^1$
together with their derivatives  not faster than any power of $|\vec x|$.

By analogy with the Schr\"odinger equation, we define the
semiclassical concentrated for the Pauli equation:
\begin{Def} The state $\Psi$ will be  called
semiclassically concentrated on the phase trajectory
$z(t)=(\vec p(t),\vec x(t))$ of the class ${\Bbb C}
{\Bbb S}_D(N)={\Bbb C}{\Bbb S}_D(N,z(t),\h)$\  $(\Psi\in{\Bbb C}
{\Bbb S}_D(N,z(t),\h))$ if
$$
\hat L_D\Psi=0;
\leqno{\rm (i)}
$$
{\rm (ii)} for the wave function $\Psi(\vec x,t,\h)$ in the
$x$-representation and for the wave function $\tilde\Psi(\vec p,t,\h)$
in the $p$-representation there are generalized limits{\rm :}
\begin{eqnarray}
&&\lim\limits_{\h\to0}|\Psi(\vec x,t,\h)|^2=\delta(\vec x-\vec x(t)),\\
&&\lim\limits_{\h\to0}|\Psi(\vec p,t,\h)|^2=\delta(\vec p-\vec p(t));
\end{eqnarray}
{\rm (iii)} for $\h\in [0,1[$ there exist quantum momenta
$^{(j)}\Delta^{{\rm cl}(k)}_{\alpha,\beta}$, $|\alpha|+|\beta|=k$,
$0\leqslant k\leqslant N$, $\alpha,\beta\in{\Bbb Z}^3_+$,
$|\alpha|=\sum\limits_{k=1}^3\alpha_k$, $j=\overline{1,16}$.
\end{Def}
Here
\begin{equation}
^{(j)}\Delta^{{\rm cl}(k)}_{\alpha,\beta}=\langle\Psi|\Gamma_j
\hat\Delta^{{\rm cl}(k)}_{\alpha,\beta}|\Psi\rangle_D,
\end{equation}
where $\Delta^{{\rm cl}(k)}_{\alpha,\beta}$ is defined in (I.0.13),
$\langle~|~\rangle_D$ is the scalar product
\begin{equation}
\langle\Psi_1|\Psi_2\rangle_D=\int d^3x\,\Psi_1^+(\vec x,t,\h)
\Psi_2(\vec x,t,\h),
\end{equation}
and $\Gamma_j$ is a certain basis in the space of Dirac matrices.

By analogy with the nonrelativistic case we have
\begin{theo} If $\Psi(t)$ is a semiclassically concentrated state
of the class ${\Bbb C}{\Bbb S}_D(z(t),N)$, then $\vec x(t)$ and
$\vec p(t)$ are solutions of the classical Hamilton system with
Hamiltonian $\lambda^{(+)}(\vec p,\vec x,t)$ or $\lambda^{(-)}(\vec p,\vec
x,t)$, where
$$
\lambda^{(\pm)}(\vec p,\vec x,t)=e\Phi(\vec x,t)\pm\ve(\vec p,\vec x,t),\qquad
\ve(\vec p,\vec x,t)=\sqrt{c^2\vec\CP^2+m_0^2c^4}.
$$
\end{theo}
{\bf Proof}.

1. We consider the spectral properties of the principal symbol of
the Hamiltonian $\hat\CH_D$, i.e., of the matrix $\CH_0(\vec p,\vec x,t)$
(2.2), $\vec p\in{\Bbb R}_p^3$, $\vec x \in{\Bbb R}_x^3$. The equation
$$
\det\|\CH_0(\vec p,\vec x,t)-\lambda{\bf I}_{4\times 4}\|=0
$$
has [28] (see Property 2, Appendix A) two eigenvalues $\lambda$ of
multiplicity 2 for all   $\vec p\in{\Bbb R}_p^3$,
$\vec x\in{\Bbb R}_x^3$, $t\in{\Bbb R}^1$:
\begin{equation}
\begin{array}{c}
\lambda^{(\pm)}(\vec p,\vec x,t)=e\Phi(\vec x,t)\pm\ve(\vec p,\vec x,t)\\
\ve(\vec p,\vec x,t) = \sqrt {c^2\vec\CP^2 + m_0^2 c^4},\quad
\vec\CP = \vec p-\frac e c \vec\CA(\vec x,t)\end{array}
\end{equation}
We combine the eigenvectors $f^\pm_j (\vec p,\vec x,t)$, $j=1,2$,
corresponding to $\lambda^{(\pm)} (\vec p,\vec x,t)$, into  $4\times2$
Pauli matrices $\Pi_\pm(\vec p,\vec x,t)$:
$$
\CH_0(\vec p,\vec x,t)\Pi_\pm(\vec p,\vec x,t) = \lambda^{(\pm)}
(\vec p,\vec x,t)\Pi_\pm (\vec p,\vec x,t),
$$
\begin{equation}
\Pi_+ (\vec p,\vec x,t)=(2\ve(\ve+m_0c^2))^{-1/2}
\left(\begin{array}{c}\ve+m_0 c^2\\ c\sp
\end{array}\right),
\end{equation}
$$
\Pi_-(\vec p,\vec x,t) = {(2\ve(\ve + m_0 c^2))}^{-1/2}
\left(\begin{array}{c} c\sp \\ -m_0 c^2-\ve
\end{array}\right).
$$

2. We expand the function  $\Psi$ with respect to eigenvectors (on
the trajectory $\vec p(t)$, $\vec x(t)$) of
the principal symbol of the Hamiltonian:
\begin{equation}
\Psi(\vec x,t,\h)=N_\h\bigl(\Pi_+(t)J^{(+)}(\vec x,t,\h)+\Pi_-(t)
J^{(-)}(\vec x,t,\h)\bigr),
\end{equation}
where $\Pi_\pm(t)=\Pi_\pm\bigl(\vec p(t),\vec x(t),t\bigr)$, $N_\h=\rm const$.
The spinors $J^{(\pm)}$ are represented as follows
\begin{equation}
J^{(\pm)}(\vec x,t,\h)=\biggl(\begin{array}{l}Q_1^{(\pm)}(\vec x,t,\h)
\,\exp\bigl( i\Xi_1^{(\pm)}(\vec x,t,\h)\bigr)\\
Q_2^{(\pm)}(\vec x,t,\h)\,\exp\bigl( i\Xi_2^{(\pm)}
(\vec x,t,\h)\bigr)\end{array}\biggr),
\end{equation}
where $\Xi_k^{(\pm)}$, $Q_k^{(\pm)}$, $k=1,2$, are real functions.
The condition (2.3), (2.4) implies that, without loss of
generality, one can set
\begin{equation}
Q_k^{(\pm)}(\vec x,t,\h)=\biggl(\prod_{l=1}^3a_l(\h)\biggr)^{-1/2}
\rho_k^{(\pm)}(\vec\xi,t,\h), \qquad \xi_j=\Delta x_j/a_j(\h),
\end{equation}
where $a_j(\h)$ are certain nonnegative functions of $\h$, such
that
\begin{equation}
\lim_{\h\to0} a_j(\h)=\lim_{\h\to0}{\h\over a_j(\h)}=0.
\end{equation}

Since there exist momenta $^{(j)}\Delta^{{\rm cl}(0)}_{0;0}$,
the following limits are exist:
$$
\lim_{\h\to0}\bigl(\Xi_k^{(\pm)}(\vec\xi,t,\h)-
\Xi_l^{(\pm)}(\vec\xi,t,\h)\bigr), \qquad
\lim_{\h\to0}\bigl(\Xi_k^{(\mp)}(\vec\xi,t,\h)-
\Xi_l^{(\mp)}(\vec\xi,t,\h)\bigr),
$$
$k,l=1,2$. Hence,
\begin{equation}
\Xi_k^{(\pm)}(\vec\xi,t,\h)=\Phi_0(\vec\xi,t,\h)+
\Phi_k^{(\pm)}(\vec\xi,t,\h), \qquad k=1,2,
\end{equation}
where $\Phi_k^{(\pm)}(\vec\xi,t,\h)$ regularly depend on  $\h$.

Similarly, we get for the wave function in the
$p$-representation:
\begin{equation}
\tilde J^{(\pm)}(\vec p,t,\h)=\biggl(\prod_{l=1}^3a_l(\h)\biggr)^{-1/2}
\biggl(\begin{array}{l}\tilde\rho_1^{(\pm)}(\vec\eta,t,\h)
\,\exp\bigl(- i\tilde\Xi_1^{(\pm)}(\vec\eta,t,\h)\bigr)\\
\tilde\rho_2^{(\pm)}(\vec\eta,t,\h)\,\exp\bigl(- i\tilde\Xi_2^{(\pm)}
(\vec\eta,t,\h)\bigr)\end{array}\biggr),
\end{equation}
$$
\eta_j=\Delta p_j/b_j(\h),\qquad \lim_{\h\to0}b_j(\h)=\lim_{\h\to0}{\h\over
b_j(\h)}=0,\qquad j=\overline{1,3},
$$
and
\begin{equation}
\tilde\Xi{}_k^{(\pm)}(\vec\eta,t,\h)=\tilde\Phi_0(\vec\eta,t,\h)+
\tilde\Phi_k^{(\pm)}(\vec\eta,t,\h), \qquad k=1,2,
\end{equation}
where $\Phi_k^{(\pm)}(\vec\eta,t,\h)$ regularly depend on $\h$.

Precisely as in the nonrelativistic case, the conditions
$$
\begin{array}{c}{}\|\Delta x_j\Psi\|=\|\Delta\hat x_j\tilde\Psi\|\sim a_j(\h),\\
{}\|\Delta\hat p_j\Psi\|=\|\Delta p_j\tilde\Psi\|\sim b_j(\h),\end{array}
\quad j=1,2,3,
$$
implies
\begin{equation}
0\leqslant\lim_{\h\to0}{\h\over a_j(\h)b_j(\h)}<\infty,\qquad j=1,2,3,
\end{equation}
as well as
\begin{equation}
\begin{array}{c}\Phi_0(\vec\xi,t,\h)=\Phi_0(t,\h)+\bigl(\langle\vec p(t),
\Delta\vec x\rangle+S(\vec\xi,t,\h)\bigr)\bigm/\h,\\
\tilde\Phi_0(\vec\eta,t,\h)=\tilde\Phi_0(t,\h)+\bigl(\langle\vec x(t),
\Delta\vec p\rangle+\tilde S(\vec\eta,t,\h)\bigr)\bigm/\h,\end{array}
\end{equation}
and
$$
S(\vec\xi,t,\h)\sim\tilde S(\vec\eta,t,\h)\sim C(\h),
$$
$$
C(\h)=\min\,\{a_1(\h)b_1(\h),\,a_2(\h)b_2(\h),\,a_3(\h)b_3(\h)\}.
$$

3. We find the limit (as $\h\to0$) of the averaged value of the operator
$\hat L_D$ with respect to the state (2.9), (2.10), (2.13), (2.17).
By  (2.1), (2.2), we get
\begin{eqnarray}
0\equiv\biggl(\lim_{\h\to0}\h{ d\Phi_0(t,\h)\over dt}-\langle\vec p(t),
\dot{\vec x}(t)\rangle+\lambda^{(+)}(t)\biggr)\lim_{\h\to0}\|J^{(+)}
(\vec x,t,\h)\|^2+{}\cr
{}+\biggl(\lim_{\h\to0}\h{ d\Phi_0(t,\h)\over dt}-\langle\vec p(t),
\dot{\vec x}(t)\rangle+\lambda^{(-)}(t)\biggr)\lim_{\h\to0}\|J^{(-)}(\vec x,t,\h)\|^2.
\end{eqnarray}
The expressions in parentheses in (2.18) cannot vanish simultaneously,
thus,
\begin{eqnarray}
&\Phi_0(t,\h)=\displaystyle{1\over\h}S_0(t,\h),\cr
&S_0(t,0)=\displaystyle\int\limits_0^t\bigl(\langle\vec p(t),\dot{\vec x}(t)\rangle-
\lambda^{(\pm)}(t)\bigr) dt,\\
&\lim\limits_{\h\to0}\|J^{(\mp)}(\vec x,t,\h)\|=0.\nonumber
\end{eqnarray}

4. For definiteness, we put
\begin{equation}
S_0(t,0)=\int\limits_0^t\bigl(\langle\vec p(t),\dot{\vec x}(t)\rangle-
\lambda^{(+)}(t)\bigr) dt,\quad\lim_{\h\to0}\|J^{(-)}(\vec x,t,\h)\|=0.
\end{equation}
By (2.3), $\lim\limits_{\h\to0}\|J^{(+)}(\vec x,t,\h)\|=1$. Then,
averaging the operator $\hat L_D$ with respect to the state (2.19)
in the higher orders of  $\h$, we obtain
\begin{eqnarray}
0&\equiv&\biggl({ dS(t,\h)\over dt}+\lambda^{(+)}(t)-\langle\dot{\vec x}
(t),\vec p(t)\rangle\biggr){}^{(+)}_{(+)}\sigma^{{\rm cl}(0)}(t,\h)+{}\cr
&+&\sum_{k=1}^3\biggl\{{}^{(+)}_{(+)}\sigma_{\xi_k}^{{\rm cl}(1)}
(t,\h)a_k(\h)\biggl[\dot p_k-e\Phi_{x_k}-
{c^2\over\ve(t)}\big\langle\vec\CP,\bigl(-
{e\over c}\vec\CA_{,x_k}\bigr)\big\rangle\biggr]+{}\cr
&+&\,^{(+)}_{(+)}\sigma_{\eta_k}^{{\rm cl}(1)}(t,\h)b_k(\h)
\biggl[\dot x_k-{c^2\over\ve(t)}\CP_k,\biggr]\biggr\}+\dots~.
\end{eqnarray}
Here
\begin{eqnarray}
&&^{(+)}_{(+)}\sigma^{{\rm cl}(0)}(t,\h)=\|J^{(+)}(\vec x,t,\h)\|,\cr
&&^{(+)}_{(+)}\sigma_{\xi_k}^{{\rm cl}(1)}(t,\h)={1\over a_k(\h)}\int
\bigl(J^{(+)}(\vec x,t,\h)\bigr)^+\Delta x_kJ^{(+)}(\vec x,t,\h) d^3x,\\
&&^{(+)}_{(+)}\sigma_{\eta_k}^{{\rm cl}(1)}(t,\h)={1\over b_k(\h)}\int
\bigl(J^{(+)}(\vec x,t,\h)\bigr)^+\Delta\hat p_kJ^{(+)}(\vec x,t,\h) d^3x.\nonumber
\end{eqnarray}
We took into account that, under the conditions  (2.20), the
order (with respect to  $\h$) of the functions
$^{(-)}_{(+)}\sigma^{{\rm cl}(1)}$, $^{(+)}_{(-)}\sigma^{{\rm
cl}(1)}$, and $^{(-)}_{(-)}\sigma^{{\rm cl}(1)}$ is less than that of the
function (2.22). Since the functions
$^{(+)}_{(+)}\sigma_{\xi_k}^{{\rm cl}(1)}(t,\h)$ and
$^{(+)}_{(+)}\sigma_{\eta_k}^{{\rm cl}(1)}(t,\h)$
are independent, Eq. (2.21) yields
\begin{equation}
\dot{\vec p}=-\lambda^{(+)}_{\vec x}(\vec p,\vec x,t), \qquad
\dot{\vec x}=\lambda^{(+)}_{\vec p}(\vec p,\vec x,t).
\end{equation}
If in (2.20) we replace $(+)\to(-)$ and $(-)\to(+)$, then in (2.23) we get
$\lambda^{(+)}\to\lambda^{(-)}$. Thus Theorem 1 is proved.
\begin{theo} If $\Psi(t)$ is a semiclassically concerntrated state
of the class ${\Bbb C}{\Bbb S}_D(z(t),N),$  then the mean values of
the quantum-mechanical spin operator, i.e., of the Bargmann
polarization pseudovector $\hat S^\mu$ {\rm [19]}, are
{\rm(}in the limit as  $\h\to0${\rm)} solutions of the
classical relativistic equations of spin motion {\rm(}the
Bargmann--Mishel--Telegdi{\rm) [20]}.
\end{theo}
{\bf Proof}. For definiteness, we assume that $\vec x(t)$
and $\vec p(t)$ satisfy the Hamilton system (2.23) and
assumption  (2.20) holds. Denote
\begin{equation}
\vec\zeta(t)=\lim_{\h\to0}\int\bigl(J^{(+)}(\vec x,t,\h)\bigr)^+\vec\sigma
J^{(+)}(\vec x,t,\h) d\vec x,
\end{equation}
\begin{equation}
a^\mu(t)=\lim_{\h\to0}\langle\Psi(\vec x,t,\h)|\hat S^\mu|
\Psi(\vec x,t,\h)\rangle_D,
\end{equation}
where
\begin{equation}
\hat S^\mu=(\hat S_0,\hat{\vec S}),\qquad \hat S_0={1\over m_0c}\langle
\vec\Sigma,\hat{\vec\CP}\rangle,\qquad \hat{\vec S}=\rho_3\vec\Sigma+
{1\over m_0c}\rho_1\hat{\vec\CP}.
\end{equation}
Then, by (2.20), we have
\begin{equation}
a_0(t)=\gamma\langle\vec\zeta(t),\vec\beta\rangle,\qquad
\vec a(t)=\vec\zeta(t)+{\gamma\vec\beta\over\ga}\langle\vec\zeta(t),
\vec\beta\rangle,\qquad a_0(t)=\langle\vec\beta,\vec a(t)\rangle.
\end{equation}
The time evolution of  mean values $a^\mu(t)$  is described by
the Heisenberg equation
\begin{equation}
\frac{ da^\mu}{ dt}=\lim_{\h\to0}\langle\Psi|\left\{
\frac{\pa\hat S^\mu}{\pa t}+{ i\over\h}[\hat\CH_D,\hat S^\mu]\right\}
|\Psi\rangle.
\end{equation}

Commuting the operator  $\hat S^\mu$ with $\hat\CH_D$, we get
(see, for example, [29] and Appendix B):
\begin{equation}
\begin{array}{l}\displaystyle\frac{\pa\hat S_0}{\pa t}+\displaystyle\frac{ i}
{\h}[\hat\CH_D,\hat S_0]=\displaystyle\frac{e}{m_0c}\langle\vec\Sigma,\vec E\rangle-\\
\qquad\qquad-\displaystyle\frac{e_0(g-2)}{2(m_0c^2)^2}(\rho_2\langle\vec\Sigma,
\vec E\times\hat{\vec\CP}\rangle+\rho_3\langle\vec\Sigma,\vec H\times
\hat{\vec\CP}\rangle)+O(\h),\\
\displaystyle\frac{\pa\hat{\vec S}}{\pa t}+\displaystyle\frac{ i}{\h}
[\hat\CH_D,\hat{\vec S}]=\displaystyle\frac{e}{m_0c}(\rho_1\vec E+\vec H
\times\vec\Sigma)-\\
\qquad-\displaystyle\frac{e_0(g-2)}{2m_0c^2}(\rho_1\vec E+\vec H\times\vec\Sigma-
\rho_3\langle\vec\Sigma,\vec E\rangle\displaystyle\frac{\hat{\vec\CP}}
{m_0c}+\displaystyle\frac{\hat{\vec\CP}}{m_0c}\rho_3\langle\vec\Sigma,
\vec H\rangle)+O(\h).\end{array}
\end{equation}
Substituting  (2.29) into (2.28) and taking into account
(A.8)--(A.14), we obtain (see Appendix B)
\begin{equation}
\frac{ d\vec a}{ dt}=\frac{ge}{2m_0c\gamma}(\vec E\langle
\vec\beta,\vec a\rangle+\vec a\times\vec H)+\frac{(g-2)e\gamma}{2m_0c}
\vec\beta(\langle\vec\beta,\vec a\times\vec H\rangle+\langle\vec a,
\vec\beta\rangle\langle\vec\beta,\vec E\rangle-\langle\vec a,\vec E\rangle).
\end{equation}
Passing from  (2.30) to the equation for $a^\mu$, and taking
into account (2.27), we get
\begin{equation}
\frac{ da^\mu}{ d\tau}=\frac{eg}{2m_0c}F^{\mu\nu}a_\nu+
\frac{(g-2)e}{2m_0c^3}\dot x^\mu\dot x_\nu F^{\nu\alpha}a_\alpha,
\end{equation}
where $F^{\mu\nu}$ is   the tensor of electromagnetic field, $\tau$
is the proper time. Equation (2.31) is the Bargmann--Michel--Telegdi equation.
Thus the theorem is proved.

\section{Passage to the two-component theory}

Each eigenvalue $\lambda^{(\pm)}(\vec p,\vec x,t)$ is associated with
its own Hamilton system:
\begin{eqnarray}
&\dot{\vec p}_\pm=-\lambda_{\vec x}^{(\pm)}(\vec p_\pm,
\vec x_\pm,t),  \quad\vec x_\pm (0,z_0) = \vec x_0,\quad
\vec p_\pm (0,z_0) = \vec p_0,\cr
&\dot{\vec x}_\pm=\lambda_{\vec p}^{(\pm)}(\vec p_\pm,
\vec x_\pm,t), \quad z_0 = (\vec p_0,\vec x_0) \in {R_{px}^6},\\
&\lambda_{\vec x}^{(\pm)}= e{\Phi}_{\vec x}(\vec x,t) -
\langle\vec\beta^\pm,\vec\CA_{\vec x}(\vec x,t)\rangle, \qquad
\lambda_{\vec p}^{(\pm)}= \pm\frac{c^2\vec\CP}{\ve} =c\vec\beta^\pm\nonumber
\end{eqnarray}
and the corresponding variational system
\begin{equation}
\left\{\begin{array}{l}\dot{\vec W}_{\pm}=-\lambda_{xp}^{(\pm)}(t)\vec W_{\pm}-
\lambda_{xx}^{(\pm)}(t)\vec Z_\pm,\\
\dot{\vec Z}_\pm=\lambda_{pp}^{(\pm)}(t)\vec W_\pm+\lambda_{px}^{(\pm)}(t)
\vec Z_\pm.\end{array}\right.
\end{equation}
The initial conditions for  system (3.2) are chosen similarly to the
scalar case (see (I.1.5)). Here we give  the explicit form of
$3\times3$-matrices in (3.2)
\begin{eqnarray*}
&&\lambda_{xx}^{(\pm)}(\vec p,\vec x,t) = \big\|\big(e(\Phi_{x_k x_j} -
\langle\vec\beta^\pm ,\vec\CA_{x_k x_j}\rangle) -
\frac{e^2}{c^2}\langle\vec\CA_{x_k},\lambda_{pp}^{(\pm)}
\vec\CA_{x_j}\rangle\big)\big\|,\\
&&\lambda_{px}^{(\pm)}(\vec p,\vec x,t)={(\lambda_{xp}^{(\pm)} (\vec p,\vec x,t))}^t
=-\frac {e}{c}\lambda_{pp}^{(\pm)}(\vec p,\vec x,t)\|\CA_{x_j}^k(\vec x,t)\|,\\
&&\lambda_{pp}^{(\pm)}(\vec p,\vec x,t) = \pm\frac{c^2}{\ve}\|(\delta_{jk} -
\beta_k^\pm\beta_j^\pm)\|,
\end{eqnarray*}
calculated at the points $z^\pm (t,z_0)$ of the phase trajectory.

From here on we restrict ourselves to solutions of ``positive frequency''
(corresponding to the eigenvalue $\lambda^{(+)}(\vec p,\vec x,t)$). Further,
in functions $\vec x_+,\dots$ related to $\lambda^{(+)}$, we shall omit the
index $+$ in all cases where it does not lead to misunderstanding.
The solutions corresponding to $\lambda^{(-)}(\vec p,\vec x,t)$ (of
``negative frequency'') are obtained by the substitution:
$\lambda^{(+)}\to \lambda^{(-)}$, $z^+(t,z_0)\to z^-(t,z_0)$,
$\Pi_\pm\to \Pi_\mp$, and $(\vec Z_+,\vec W_+)\to(\vec Z_-,\vec W_-)$.

Let us quantize the classical system (3.1) by the method of
complex germ, i.e., with an arbitrary (but fixed) trajectory
of a classical particle $z(t,z_0)$ we associate the complete set of
functions of the form:
\begin{equation}
|\nu,t\rangle=\prod_{k=1}^3\frac{1}{\sqrt{\nu_k !}}(\hat a_k^+(t))
^{\nu_k}|0,t\rangle,
\end{equation}
where
$$
|0,t\rangle = N_\h(J(t,z_0))^{-1/2} \exp\big[
\frac {i}{\h} S(\vec x,t)\big],
$$
$$
S(\vec x,t) = \int\limits_0^t (\langle\dot {\vec x} (t),\vec p (t)\rangle -
\lambda^{(+)}(t))\,dt + \langle\vec p (t),\Delta \vec x \rangle +
\frac {1}{2}\langle\Delta\vec x,Q(t) \Delta \vec x \rangle,
$$
$N_0(\h)$, $J(t)$, $\hat a^+(t)$ are defined in (I.2.11), (I.2.12),
respectively. It is easy to see (item 2.2 Part I) that the functions
(3.3) satisfy the equation of Schr\"odinger type:
\begin{eqnarray*}
&\qquad(-i\h\pa_t + \hat\lambda)|\nu,t\rangle = 0,\\
&\hat {\lambda} = \lambda^{(+)}(t) + \langle\dot {\vec p}(t),\Delta \vec x
\rangle-\langle\dot {\vec x} (t),\Delta \hat{\vec p}\rangle
+ \displaystyle\frac {1}{2}\big[\langle\Delta \vec x,
\lambda_{xx}^{(+)}(t)\Delta \vec x \rangle +
\langle\Delta \vec x,\lambda_{xp}^{(+)}(t)\Delta\hat{\vec p}\rangle \\
&+\langle\Delta \hat{\vec p},\lambda_{px}^{(+)}(t)\Delta\vec x\rangle+
\langle\Delta\hat{\vec p},\lambda_{pp}^{(+)}(t)\Delta\hat{\vec p}
\rangle\big] = \lambda^{(+)}(t)+\hat\delta^1\lambda^{(+)}(t) +
\displaystyle\frac{1}{2}\hat\delta^2\lambda^{(+)}(t).
\end{eqnarray*}

The eigenvectors (2.8) of the matrix $\CH_0(\vec p,\vec x,t)$
form an orthonormalized basis in ${\Bbb C}{}^4$ (see Property 11 in
Appendix A):
\begin{eqnarray*}
&&\Pi_\pm^+(\vec p,\vec x,t)\Pi_\pm(\vec p,\vec x,t)=
{\bold I}_{2\times2},\\
&&\Pi_\pm^+ (\vec p,\vec x,t) \Pi_\mp (\vec p,\vec x,t) = 0,\\[2pt]
&&\sum\limits_{k=\pm}{\Pi}_k(\vec p,\vec x,t){\Pi}_k^+(\vec p,\vec x,t) =
{\bold I}_{4\times 4},
\end{eqnarray*}
and the system of scalar functions (3.3) is complete in $\CP^t_\h({\Bbb R}^3,
{\Bbb C})$:
$$
\langle t,\nu'|\nu,t\rangle =\delta_{\nu'\nu};\qquad \sum_{|\nu|=0}^\infty
|\nu,t\rangle  \langle t,\nu| =1,
$$
and
\begin{eqnarray}
&\CP_\h^t({\Bbb R}^3,{\Bbb C}) = \bigg\{f, f=\exp{\displaystyle\frac i {\h}
(S_0(t)+\langle\vec p(t),\Delta \vec{x}\rangle)} \phi\big( \frac {\Delta \vec{x}}
{\sqrt{\h}}, t, \h\big),\quad \phi(\vec\zeta,t,\h)\in {\Bbb S}\bigg\},  \\
&S_0(t)=\int\limits_0^t (\langle \vec p (t), \dot{\vec x} (t)\rangle -\lambda^{(+)}
(t))dt,\nonumber
\end{eqnarray}
where $\phi(\vec\zeta,t,\h)$ is a smooth function in $t \in [0,T]$
regularly depends on $\h$, and ${\Bbb S}$ is a Schwartz space with
respect $\vec\zeta \in {\Bbb R}^3$. The solution of equation (2.1)
will be sought in the form
\begin{equation}
^{(+)}\Psi(\vec x,t,\h)=\Psi(\vec x,t,\h) = ({\Pi}_+ (t),{\Pi}_- (t))\left(
\begin{array}{c} {\cal U}(\vec x,t,\h) \\ {\cal V} (\vec x,t,\h)
\end{array} \right)=(\Pi_+{\cal U}+\Pi_-{\cal V}),
\end{equation}
where the matrices (2.8) are calculated at the point $z(t,z_0)$ and the
unknown two-component spinors ${\cal U}(\vec x,t,\h)\in\CP_\h^t({\Bbb R}^3,
{\Bbb C}^2)$ and ${\cal V}(\vec x,t,\h)\in\CP_\h^t({\Bbb R}^3,{\Bbb C}^2)$
must be determined.
\begin{rem} Note, that if $\Psi\in\CP_\h^t({\Bbb R}^3,{\Bbb C}^4)$ (3.5) the
solution of the Dirac equation than $\Psi$ is semiclassically concentrated
states of class ${\Bbb S}{\Bbb C}_D\big(z(t),\infty\big)$.
\end{rem}

We substitute the function (3.5) into (2.1) and expand the
obtained expressions with respect to the eigenvectors (2.8) taking into
account the relations (see Appendix A)
\begin{eqnarray*}
&&\dot\Pi_\pm(t) = \frac{i}{2}\Pi_\pm(t)\frac{\sbb}{\ga}\mp\frac{\gamma}{2}
\Pi_\mp(t)\biggl(\sb\frac{\bdb}{\ga}+\gamma^{-1}\sdb\biggr),\\
&&\rho_1\sp\Pi_\pm(t)=\pm\bp\Pi_\pm(t)+\Pi_\mp(t)\biggl(\sb\frac{\bp}{\ga}-
\sp\biggr),\\
&&\rho_2\langle\vec\Sigma,\vec E \rangle\Pi_\pm (t) =-\Pi_\pm(t)\sbe\mp
i\Pi_\mp(t)\biggl(\sb\frac{\bve}{\ga} +\gamma^{-1}\sve\biggr),\\
&&\rho_3\langle\vec\Sigma,\vec H\rangle\Pi_\pm(t) =\mp\Pi_\pm(t)\biggl(\sb
\frac{\bvh}{\ga}-\svh\biggr)+\bvh\Pi_\mp(t).
\end{eqnarray*}
Here $c\vec\beta=\dot{\vec x}(t,z_0)$, $\gamma^{-1}=\sqrt{1-\beta^2}$.
As the result we get
\begin{eqnarray}
\lefteqn{(-i\h\pa_t + \hat\CH)\Psi=\Pi_+ \bigg\{\Big[-i\h\pa_t+\lambda^{(+)}(t)
+ e\Delta\Phi+c\langle\vec \beta,\Delta \hat{\vec\CP}\rangle+{}}\cr
&&{}+\frac{\h}{2}\frac{\sbb}{\ga}-\eg\Big(\svh
- \sbe-\sb\frac{\bvh}{\ga}\Big)\Big]{\cal U}(\vec x,t)+{}\cr
&&{}+\Big[-\frac{i\h\gamma}{2}\Big(\sb\frac{\bdb}{\ga}+\gamma^{-1}\sdb
\Big) +c\Big(\sb\frac{\langle\vec\beta,\Delta\hat{\vec\CP}
\rangle}{\ga}-\langle\vec\sigma,\Delta\hat{\vec\CP}\rangle\Big)-{}\cr
&&{}- \eg\Big(i\sb\frac{\bve}{\ga} +i\gamma^{-1}\sve +\bvh\Big)
\Big]{\cal V}(\vec x,t)\bigg\}+{}\cr
&&{}+\Pi_-\bigg\{\Big[-i\h\pa_t+\lambda^{(-)}(t)+e\Delta\Phi -c\langle\vec\beta,
\Delta\hat{\vec\CP}\rangle+\frac {\h}{2}\frac{\sbb}{\ga}-{}\cr
&&{}-\eg\Big(-\sbe-\svh +\sb\frac{\bvh}{\ga}\Big)\Big]{\cal V}(\vec x,t)+{}\cr
&&{}+\Big[\frac{i\h\gamma}{2}\Big(\sb\frac{\bdb}{\ga} +\gamma^{-1}\sdb\Big)
+c\Big(\sb\frac{\langle\vec\beta,\Delta\hat{\vec\CP}\rangle}{\ga}
-\langle\vec\sigma,\Delta\hat{\vec \CP}\rangle\Big)-{}\cr
&&{}-\eg\Big(-i\sb\frac{\bve}{\ga} -i\gamma^{-1}\sve+\bvh\Big)
\Big]{\cal U}(\vec x,t)\bigg\},
\end{eqnarray}
where $\Delta\Phi = \Phi(\vec x,t)-\Phi(\vec x(t,z_0),t)$,
$\Delta\hat{\vec\CP} =\hat{\vec\CP}-\vec\CP(t)$.

We transform the obtained equation by expanding the expressions
in this equation into a Taylor series in operators $\Delta\vec x$ and
$\Delta\hat{\vec p}$ and taking into account that in the class of
functions $\CP_\h^t$ we have
$$
\Delta \vec x =\hat O(\sqrt{\h}), \quad
\Delta\hat{\vec p} =\hat O(\sqrt{\h}), \quad
(-i\h\pa_t+\hat\lambda)=\hat O(\h),
$$
as $\h\to0$ (see (I.4.4)). Denote $\hat{\vec\CP}_1=
\Delta\hat{\vec p}-(e/c)d^1\vec\CA(t)$,
\begin{eqnarray}
\hat R^{(N)}&=&\frac{i\h\gamma}{2}\big(\sb\frac{\bdb}{\ga} +
\frac{\sdb}{\gamma}\big) +c\big(\sb\frac{\langle\vec\beta,\hat{\vec\CP}_1\rangle}
{\ga} -\langle\vec\sigma,\hat{\vec\CP}_1\rangle\big)-{}\cr
&-&\sum_{k=2}^{N+2}\frac {e}{k!}\Big(\sb\frac{\langle\vec\beta,d^k\vec\CA\rangle}{\ga} -
\langle\vec\sigma,d^k\vec\CA\rangle \Big)-\eg\sum_{k=0}^N \frac{1}{k!}
\Big(- i\sb\frac{\langle\vec\beta,d^k\vec E\rangle}{\ga}-{}\cr
&-&i\gamma^{-1} \langle\vec\sigma,d^k\vec E\rangle +
\langle\vec\beta,d^k\vec H\rangle \Big) + \hat O(\h^{(N+3)/2}),\\
\hat{\cal M}^{(N)}&=&2\ve(t) -2c\langle\vec\beta,
\vec\CP_1\rangle +(-i\h\pa_t + \hat\lambda) -\frac{1}{2}
\langle\hat{\vec\CP}_1,\lambda_{pp}^{(+)}\hat{\vec\CP}_1\rangle+{}\cr
&+&\sum_{k=3}^{N+2}\frac{e}{k!}(d^k\Phi+\langle\vec\beta,d^k\vec\CA\rangle)
+e\langle\vec\beta,d^2\vec\CA\rangle +\frac{\h}{2}\frac{\sbb}{\ga}-{}\cr
&-&\eg\sum_{k=0}^N\frac{1}{k!}\Big(-\langle\vec\sigma,\vec\beta
\times d^k\vec E\rangle -\langle\vec\sigma,d^k\vec H\rangle+\sb
\frac{\langle\vec\beta,d^k\vec H\rangle}{\ga}\Big) +\hat O(\h^{(N+3)/2}).
\end{eqnarray}
Then equation (3.6) takes the form
\begin{eqnarray*}
\lefteqn{(-i\h\pa_t + \hat\CH)\Psi^{(N)} =\Pi_+\bigg\{\Big[-i\h\pa_t + \hat\lambda -
\frac{1}{2}\langle\hat{\vec\CP}_1,\lambda_{pp}^{(+)}\hat{\vec\CP}_1\rangle+{}}\\
&&{}+\frac{\h}{2}\frac{\sbb}{\ga} + \sum_{k=3}^{N+2} \frac{e}{k!} (d^k\Phi -
\langle\vec\beta,d^k\vec \CA\rangle)-{}\\
&&-\eg\sum_{k=0}^N \Big(\langle\vec\sigma,d^k\vec H\rangle -
\langle\vec\sigma,\vec\beta\times d^k\vec E\rangle - \sb\frac
{\langle\vec\beta,d^k\vec H\rangle}{\ga}\Big)\Big]{\cal U}^{(N)}(\vec x,t) +{}\\
&&{}+({\hat R}^{(N)})^+{\cal V}^{(N)}(\vec x,t)\bigg\}+
\Pi_- \big[\hat{\cal M}^{(N)}{\cal V}^{(N)}(\vec x,t) + {\hat R}^{(N)}
{\cal U}^{(N)}(\vec x,t)\big]+ \hat O(\h^{(N+3)/2}).
\end{eqnarray*}
Since the eigenvectors of the principal symbol of the Hamiltonian
$\hat\CH_D$, which define the matrices $\Pi_+(t)$ and $\Pi_-(t)$,
are linearly independent in ${\Bbb C}{}^4$, then, by (3.5), we get
\begin{eqnarray}
\lefteqn{{\cal V}^{(N)}(\vec x,t)=-\{\hat{\cal M}^{(N)}\}^{-1}
{\hat R}^{(N)}{\cal U}^{(N)}(\vec x,t),}\\
\lefteqn{(-i\h\pa_t + \hat F^{(N)}){\cal U}^{(N)}(\vec x,t)=0,}\\
\lefteqn{\hat F^{(N)} =\hat\lambda-\frac{1}{2}\langle\hat{\vec\CP}_1,\lambda_{pp}^{(+)}
\hat{\vec\CP}_1\rangle +\sum_{k=3}^{N+2}\frac{e}{k!}(d^k\Phi -
\langle\vec\beta,d^k\vec\CA\rangle)+{}}\cr
&&{}+ \frac{\h}{2}\frac{\sbb}{\ga}-\eg\sum_{k=0}^N\frac{1}{k!}
\Big(\langle\vec\sigma,d^k\vec H\rangle -\langle\vec\sigma,\vec\beta\times
d^k\vec E\rangle-{}\cr
&&{}-\sb\frac{\langle\vec\beta,d^k \vec H\rangle}{\ga}\Big)-\{{\hat R}^{(N)}\}^+
\{\hat{\cal M}^{(N)}\}^{-1}{\hat R}^{(N)} +\hat O(\h^{(N+3)/2}).
\end{eqnarray}
The operator $({\cal M}^{(N)})^{-1}$ inverse to the operator (3.8) will
be found from (I.4.10) where we put $\hat B=\widehat{\cal M}-2\ve$,
$\hat A=2\ve$. Then
\begin{equation}
\frac{1}{2\ve}{\hat Q}^{(N)} = - \{\hat{\cal M}^{(N)}\}^{-1}
{\hat R}^{(N)} = \frac{1}{2\ve}\sum_{k=1}^{N+2} \h^{k/2}
\hat Q_k + \hat O(\h^{(N+3)/2}),
\end{equation}
where
\begin{eqnarray}
\sqrt{\h}\hat Q_1&=&c\Big(\sb\frac{\langle\vec\beta,\hat{\vec\CP}_1\rangle}
{\ga} -\langle\vec\sigma,\hat{\vec\CP}_1\rangle\Big),\\
\h\hat Q_2&=&- \frac{e}{2}\Big(\sb\frac{\langle\vec\beta,d^2\vec\CA\rangle}
{\ga}-\langle\vec\sigma,d^2\vec\CA\rangle\Big)+\frac{i\gamma\h}{2}
\Big(\gamma^{-1}\sdb+\sb\frac{\bdb}{\ga}\Big)-{}\cr
&&{}-\eg\big[i\sb\frac{\bvet}{\ga} + i\gamma^{-1}\svet-\bvht\big]
- \frac{c\sqrt{\h}}{\ve}\langle\vec\beta,\hat{\vec\CP}_1\rangle\hat Q_1,\nonumber
\end{eqnarray}
and for $k>2$
\begin{eqnarray*}
\lefteqn{\h^{k/2}\hat Q_k = -\frac{e}{k!}\Big(\sb\frac{\langle\vec\beta,
d^k\vec\CA\rangle}{\ga}-\langle\vec\sigma,d^k\vec\CA\rangle\Big)+\eg
\frac{1}{(k-2)!}\Big(\langle\vec\beta,d^k\vec H\rangle -{}}\\
&&{}- i\sb\frac{\langle\vec\beta,d^{k-2}\vec E\rangle}{\ga}-
 i\gamma^{-1}\langle\vec\sigma,d^{k-2}\vec E\rangle\Big) -\frac{c}{\ve}
\langle\vec\beta,\hat{\vec\CP}_1\rangle\hat Q_{k-1} +
\Big[e\langle\vec\beta,d^2\vec \CA\rangle +(- i\h\pa_t + \hat\lambda) -{}\\
&&{}-\frac 1 2 \langle\hat{\vec\CP}_1,\lambda_{pp}^{(+)}\hat{\vec\CP}_1\rangle+
\frac{\h}{2}\frac{\sbb}{\ga}-\eg\Big(\svht - \sb\frac{\bvh}{\ga} + {}\\
&&{}+\sbet\Big)\Big]\frac{\h^{(k-2)/2}}{2\ve}{\hat Q}_{k-2}+\sum_{n=3}^{k-1}
\Big[\frac{e}{n!}(d{}^n\Phi+\langle\vec\beta,d{}^n\vec\CA\rangle) +{}\\
&&{}+\eg\frac{1}{(n-2)!}\Big(\langle\vec\sigma,\vec\beta\times
d{}^{n-2}\vec E\rangle + \langle\vec\sigma,d{}^{n-2}\vec H\rangle-
\sb\frac{\langle\vec\beta,d{}^{n-2}\vec H\rangle}{\ga}\Big)\Big]
\frac{\h^{(k-n)/2}}{2\ve}{\hat Q}_{k-n}.
\end{eqnarray*}
By taking into account (3.8),\dots, (3.11), the operator $\hat F^{(N)}$
can be represented in the form
$$
{\hat F}^{(N)} = {\hat F}_0 + \sqrt {\h}{\hat F}_1^{(N)},\qquad
{\hat F}_1^{(0)} = 0 ,
$$
where
\begin{eqnarray}
\hat F_0 = \hat\lambda+\frac{\h}{2}\frac{\sbb}{\ga} - \frac{\h}{2}
\frac{ec}{\ve}\Big(\frac{\langle\vec\sigma,\vec\beta\times(\vec\beta\times
\vec H(t))\rangle}{\ga}+\svht\Big)+{}\cr
{}+\eg\big(\svht-\sbet-\sb\frac{\bvht}{\ga} \big),
\end{eqnarray}
and the highest terms have the form $\hat F_1^{(N)} =\hat O(\h)$ as $\h\to0$.

Thus we have
\begin{equation}
{\hat T}^{(N)}(t)=\big(\Pi_+(t)+ \frac{1}{2\ve}\Pi_-(t){\hat Q}^{(N)}(t)\big)
\end{equation}
and the problem of constructing asymptotic (up to $O(\h^{(N+1)/2})$)
positive-frequency solutions $^{(+)}\Psi(\vec x,t,\h)$ to the Dirac
equation (2.1) is reduced to solving the equation (3.10) with respect to the
two-component spinor ${\cal U}^{(N)}(\vec x,t)$. As a result we have
$$
^{(+)}\Psi(\vec x,t,\h)=\Psi(\vec x,t,\h)=\hat T^{(N)}(t){\cal U}^{(N)}
(\vec x,t).
$$

\section{Semiclassical trajectory-coherent representation
of the Dirac equation with anomalous Pauli interaction}

Now let us consider the construction of asymptotic (as $\h\to0$)
solutions of equations (3.10) in the two-component theory:
\begin{equation}
(-i\h\pa_t + \hat F_0 + \sqrt{\h}\hat F_1^{(N)}){\cal U}^{(N)}(\vec x,t,\h)
= 0.
\end{equation}
In the first approximation as $\h\to0$ (neglecting the
operator $\sqrt{\h}\hat F_1=\hat O(\h^{3/2})$ in (3.10)) we have
the following equation for the spinor ${\cal U}^{(0)}(t)$:
\begin{equation}
(-i\h\pa_t + \hat F_0){\cal U}^{(0)}(\vec x,t,\h)= 0
\end{equation}
and hence ${\cal U}^{(0)}(\vec x,t,\h)=\phi(\vec x,t,\h)u(t)$,
\begin{eqnarray}
&\displaystyle\Big\{i\frac{d}{dt} + \frac{ec}{2\ve}\Big[(1+\tilde g\gamma)
\svht-(\frac{1}{\ga} +\tilde g\gamma)\sbet -{}\cr
&{}-\displaystyle\frac{\tilde g\gamma\bvht}{\ga}\sb\Big]\Big\}u = 0,\\
&(-i\h\pa_t+\hat\lambda)\phi(\vec x,t,\h)=0,\nonumber
\end{eqnarray}
where $\tilde g=(g-2)/2$. The derivation of (4.3) was based on the following
relation [30]:
\[ \dot{\vec\beta} = \frac{ec}{\ve}(\vec E + \vec\beta \times \vec H -
\vec\beta \langle\vec\beta,\vec E\rangle). \]
Thus the spinor properties of electrons in the semiclassical
limit as $\h\to0$ are determined by complex solutions of
the linear system (4.3). Assume that at the initial time the
spinor $u(t)$ satisfies the following condition  \cite{31}:
\begin{equation}
\langle\vec\sigma,\vec\ell\rangle u(0,\zeta) = \zeta u(0,
\zeta),\quad\zeta = \pm1.
\end{equation}
By this assumption we fix that at $t=0$ the spin of a particle is
directed along an arbitrary unit vector $\vec\ell\in {\Bbb R}^3$. Then at any
instant of time the solutions $u(t,\zeta)$ of the Cauchy
problem (4.3), (4.4) form an orthonormalized basis in ${\Bbb C}{}^2$:
$u^+(t,\zeta')u(t,\zeta)=\delta_{\zeta,\zeta'}$ and, hence,
the functions
$$
{\cal U}^{(0)}_{\nu,\zeta}(\vec x,t,\h)=u(t,\zeta)|\nu,t\rangle =
|\nu,\zeta,t\rangle
$$
form the complete orthonormalized set of solutions of the equation (4.2)
$$
\langle t,\zeta',\nu'|\nu,\zeta,t\rangle=\delta _{\nu,\nu'}
\delta_{\zeta,\zeta'}.
$$
The precision of constructed solutions can be estimated
similarly to the case of the Schr\"odinger equation (see Theorem 2 in [1]).

Let us construct the following (in $\h\to0$) approximations
for  equation (4.1). In contrast to the nonrelativistic case,
the operator of ``perturbation'' $\sqrt{\h}\hat F{}_1^{(N)}$
is not self-adjoint\footnote{A similar situation takes place for
the highest nonrelativistic approximations \cite{32}.}
in $L_2(R^3,{\Bbb C}{}^2)$ and hence the function ${\cal U}^{(N)}(\vec
x,t,\h)$ cannot be considered as the wave function of electron. However,
to solve equation (4.1) by the methods of the theory  of perturbations
and then to go over to the semiclassical TC-representation, it is
sufficient that only the operator $\hat F_0$ be self-adjoint. By using
formula (I.4.10) with operators
$$
\hat A=\pa_t + \frac{i}{\h} \hat F_0,\quad
\epsilon\vec B = \frac{i}{\sqrt {\h}} \hat F{}_1^{(N)}
$$
and taking into account that
$$
\hat A^{-1}\phi(t) = \sum_{|\nu|=0}^{\infty} \sum_{\zeta=\pm1}
|\nu,\zeta,t\rangle \int\limits_0^t d\tau\langle\tau,\zeta,\nu|
\phi(\tau)\rangle,
$$
we obtain the solution of equation (4.1) with precision up to $O(\h^{(N+1)/2})$:
\begin{eqnarray}
&&{\cal U}_{\nu,\zeta}^{(N)}(\vec x,t,\h) =
\hat\CF^{(N)}|\nu,\zeta,t\rangle + O(\h^{(N+1)/2}),\cr
&&\hat\CF^{(N)} =\sum_{n=0}^N\Big(-\frac{i}{\sqrt{\h}}\Big)^n
\big(\hat\CF_1^{(N)}\big)^n,\\
&&\hat\CF_1^{(N)}\phi(t)=\sum_{|\nu'|=0}^{\infty}\sum_{\zeta'=\pm1}
|\nu',\zeta',t\rangle \int\limits_0^td\tau\langle\tau,\zeta',\nu'|
\hat F_1^{(N)}|\phi(\tau)\rangle.\nonumber
\end{eqnarray}

Now let us construct the semiclassical TC-representation for the
Dirac equation (2.1) following the scheme of constructing the
TC-representation for the Schr\"odinger operator. We introduce
the Hilbert space of vector-functions
\[ L_\h^t({\Bbb R}^3,{\Bbb C}^2) \]
with scalar product
$$
\langle\va_1 | \va_2\rangle_{L_\h^t} = \int d^3x\rho_\h^{z_0}
(\vec x,t)\va_1^+ (\vec x,t,\h)\va_2 (\vec x,t,\h),
$$
where the density of measure $\rho_\h^{z_0}(\vec x,t)$ was defined in (1.2).

We define the operator $\hat\CK_D^{(N)}(t,\h):L_\h^t\to
L_2({\Bbb R}^3,{\Bbb C}^4)$, which defines the passage to the
semiclassical TC-representation, by the formula:
\begin{equation}
\hat\CK_D^{(N)}(t,\h)\va =\hat T^{(N)}\hat\CF^{(N)}
\hat\CK_D^{(0)}(t,\h)\va,\quad\va\in L_{\h}^t({\Bbb R}^3,{\Bbb C}^2),
\end{equation}
where the operator $\hat T^{(N)}$ is defined in (3.15), $\hat\CK_D^{(0)}(t,\h)=
\hat\CK_S^{(0)}(t,\h)$, and $\hat\CK_S^{(0)}(t,\h)$ is defined in (1.4), in
which the symbol $\CH(\vec p,\vec x,t)$ must be replaced by the relativistic
Hamiltonian function $\lambda^{(+)}(\vec p,\vec x,t)$.

The operator $\hat\CK_D^{(N)}(t,\h)$  (with precision up to
$O(\h^{(N+1)/2})$) unitarily maps the space $L_\h^t({\Bbb R}^3,{\Bbb C}^2)$
into the space $L_2({\Bbb R}^3,{\Bbb C}^2)$, this means that
$$
\langle\hat\CK_D^{(N)}(t,\h)\va_1|\hat\CK_D^{(N)}(t,\h)\va_2\rangle_{L_2} =
\langle\va_1|\va_2\rangle_{L_\h^t}+O(\h^{(N+1)/2}).
$$
By direct calculations it is easy to verify that in the
semiclassical TC-representation  the Dirac equation (2.1) takes the form:
\begin{equation}
(\hat\CK_D^{(N)}(t,\h))^+(-i\h\pa_t+ \hat\CH_D)\hat\CK_D^{(N)}(t,\h)\va =
\big[\hat\pi_0 +\langle\vec\sigma,\vec{\cal D}_0(t,z_0)\rangle\big]\va+
\hat O(\h^{(N+3)/2}),
\end{equation}
where the operator $\hat\pi_0$ is given by the formula (1.7), in which
the symbol $\CH(\vec p,\vec x,t)$ must be replaced by the relativistic
Hamiltonian function $\lambda^{(+)}$ (2.7), and the vector
$\vec{\cal D}_0(t,z_0)$ is equal to
\begin{equation}
\vec{\cal D}_0(t,z_0)=\frac{\mu_0}{\gamma}\Big[(1+\tilde g\gamma)\vec H (t)-
\Big(\frac{1}{\ga}+\tilde g\gamma\Big)\vec\beta\times\vec E(t)-\frac{\tilde g\gamma}
{\ga}\vec\beta\bvht\Big],
\end{equation}
where $\mu_0 = \h e_0/(2m_0c)$ is Bohr magneton.

Thus, with precision up to $\hat O(\h^{(N+3)/2})$, the Dirac equation
in the semiclassical representation takes the form:
\begin{equation}
(\hat\pi_0+\langle\vec\sigma,\vec{\cal D}_0(t,z_0)\rangle)\va = 0.
\end{equation}
Precisely as in (1.8), equation (4.9) admits the complete
orthonormalized set of solutions of the form:
\begin{equation}
|H_{\nu},\zeta\rangle =u(t,\zeta) \prod_{k=1}^{3} \frac {1}
{\sqrt{\nu_k!}}{(\hat\La^+)}^{\nu_k} \cdot 1,
\end{equation}
where the operators of creation $\hat\La^+_k$ are given by  formulas
(1.9), in which the complex vectors $\vec W_j(t)$ and $\vec Z_j(t)$
are solutions of the variational system (3.2).

We return to the initial (Schr\"odinger) representation of the
Dirac equation (2.1) and, taking into account (4.6) and (4.9),
obtain the complete orthonormalized set of semiclassical (up to
$\hat O(\h^{(N+1)/2})$) trajectory-coherent states of electron:
\begin{equation}
\Psi_{\nu,\zeta}^{(N)}(\vec x,t,\h) = \hat\CK_D^{(N)}(t,\h)|H_\nu,\zeta\rangle.
\end{equation}
The negative-frequency semiclassical TC-states $^{(-)}\Psi_{\nu,\zeta}^{(N)}$
are given by the same formula where we change:
\begin{eqnarray*}
&\lambda^{(+)} \longrightarrow \lambda^{(-)}, \quad
\Pi_\pm (t) \longrightarrow \Pi_\mp (t), \quad
z^+(t,z_0) \longrightarrow z^-(t,z_0),\\
&(\vec W_+ (t),\vec Z_+ (t)) \longrightarrow (\vec W_- (t),\vec Z_- (t)).
\end{eqnarray*}
\begin{rem} The relations {\rm (4.6)--(4.11)} allow us to obtain the Green
function for Dirac equation in semiclassical trajectory-coherent approximation.
Analogously {\rm (I.6.4)} for the positive-frequency part of a kernel of
evolution operator for the equation {\rm (2.1)}
\begin{equation}
G_D^{(N)}(\vec x,\vec y,t,s) = \hat T^{(N)}(t)\hat\CF^{(N)}(t)
G_D^{(0)}(\vec x,\vec y,t,s)\big(\hat\CF^{(N)}(s)\big)^+\big(\hat T^{(N)}(s)\big) ^+,
\end{equation}
where
\begin{equation}
G_D^{(0)}(\vec x,\vec y,t,s)=G^{(0)}(\vec x,\vec y,t,s)\sum_{\zeta=\pm 1}
u(t,\zeta)u^+(s,\zeta),
\end{equation}
$u(t,\zeta)$ is a solution of equation {\rm (4.3)} with initial
condition {\rm (4.4)}, $G^{(0)}(\vec x,\vec y,t,s)$ was defined in
{\rm (I.A1.1)} for $\hat{\CH}_0=\hat\lambda$. The operators $\hat T^{(N)}$
and $\hat\CF^{(N)}$ were defined in {\rm (3.15)} and {\rm (4.5)}, respectively.
Analogously to the scalar case, the Green function allows to obtain
a solution of Cauchy problem for Dirac equation {\rm (2.1)}
only in the class of semiclassical trajectory-coherent states.
Nevertheless the Green function {\rm (4.12)} can be usefull for calculation
of concrete physical effects {\rm \cite{29, 33}}.
\end{rem}

\section{The relativistic analog of the Pauli equation}

The semiclassical description of a quantum particle with its
spin properties taken into account allows (with an arbitrary
precision in $\h\to 0$) to exclude the interference between the
positive-frequency and negative-frequency states (the Schr\"odinger
``zitterbewegung'' \cite{29}), i.e., in the subspace of positive-frequency
(negative-frequency) states it allows to go over to the one-particle
two-component theory, whose Hamiltonian is a self-adjoint operator for
any order of $\h\to0$.

The unitary operator $\hat{\tilde T}{}^{(N)}(\bmod \hat O(\h^{(N+3)/2}))$
of transition to the two-component theory will be sought in the form:
\begin{equation}
\hat{\tilde T}{}^{(N)} = \hat T^{(N)}\hat B^{(N)} +\hat O (\h^{(N+3)/2}),
\end{equation}
where the operator $\hat T^{(N)}$ is defined in (3.15) and the operator
$\hat B^{(N)}$ is defined by the condition that the operator
$^{(+)}\hat{\tilde\CH}{}^{(N)}$ is self-adjoint:
\begin{equation}
^{(+)}\hat{\tilde\CH}{}^{(N)} =(\hat B^{(N)})^+(-i\h\pa_t + \hat F^{(N)})
{\hat B}^{(N)}+i\h\pa_t + \hat O (\h^{(N+3)/2}).
\end{equation}
In this case the spinor
$$
\tilde{\cal U}_{\nu,\zeta}^{(N)}=\big(\hat B^{(N)}\big)^{-1}
{\cal U}_{\nu,\zeta}^{(N)}(\vec x,t)
$$
can be considered as the  wave function of the one-particle problem:
\begin{equation}
(\hat{\tilde T}{}^{(N)})^+(-i\h\pa_t +\hat\CH_D)\hat{\tilde T}{}^{(N)}
\tilde{\cal U}^{(N)} =(-i\h\pa_t +{}^{(+)}\hat{\tilde\CH}{}^{(N)})
\tilde{\cal U}^{(N)} + \hat O(\h^{(N+3)/2}).
\end{equation}

Let us consider the construction of the operator $\hat B^{(N)}$ in the case
$N=2$ more precisely. For this purpose, we represent the not
self-adjoint part of the operator $\hat F_1^{(2)}$ as follows:
\begin{eqnarray*}
&\displaystyle\frac{\sqrt{\h}}{2\ve}\hat Q_1(-i\h\pa_t + \hat F_0)
\displaystyle\frac{\sqrt{\h}}{2\ve}\hat Q_1 =\displaystyle\frac{\h}{2}
\big\{(-i\h\pa_t+\hat F_0)\big(\displaystyle\frac{1}{2\ve}\hat Q_1\big)^2 +
\big(\displaystyle\frac{1}{2\ve}\hat Q_1\big)^2(-i\h\pa_t + \hat F_0)\big\}+{} \\
&{}+\displaystyle\frac{\h}{2}\big\{\big[\big(\displaystyle\frac{1}{2\ve}
\hat Q_1\big),(-i\h\pa_t + \hat F_0)\big]_-\displaystyle\frac{1}{2\ve}
\hat Q_1 +\displaystyle\frac{1}{2\ve}\hat Q_1\big[(-i\h\pa_t + \hat F_0),
\big(\displaystyle\frac{1}{2\ve}\hat Q_1\big)\big]_-\big\},\\
&[\hat A,\hat B]_- = \hat A\hat B - \hat B\hat A.
\end{eqnarray*}
If in Eq. (5.1) we choose the operator $\hat B^{(2)}$ as
\begin{equation}
\hat B^{(2)} = 1-\frac{\h}{2}\big(\frac{1}{2\ve}\hat Q_1\big)^2,
\end{equation}
then it can be easily verified that the conditions imposed on
the operator $\hat{\tilde T}{}^{(N)}$ are satisfied. In this case the spinor
$\tilde{\cal U}^{(2)}_{\nu,\zeta}$ takes the form:
\begin{eqnarray}
\tilde{\cal U}_{\nu,\zeta}^{(2)}(\vec x,t,\h)&=&(\hat B^{(2)})^{-1}
{\cal U}_{\nu,\zeta}^{(2)}(\vec x,t,\h)={} \cr
&=&\big[ 1-i\sqrt{\h}\hat{\tilde\CF}_1 -i\h\hat{\tilde\CF}_2 -
\h\hat{\tilde\CF}{}_1^2\big]|\nu,\zeta,t\rangle = \hat{\tilde\CF}{}^{(2)}
|\nu,\zeta,t\rangle,
\end{eqnarray}
where
\begin{eqnarray*}
\sqrt{\h}\hat{\tilde\CF}_1\va(t)&=&\displaystyle\frac{1}{\h}\sum_{|\nu'|=0,
\zeta'=\pm1}^{\infty} | \nu',\zeta',t\rangle \int\limits_0^t d\tau\langle\tau,
\zeta',\nu'|{\hat F_1}^{(1)}|\va(\tau)\rangle,\\
\h\hat{\tilde\CF}_2\va(t)&=&\displaystyle\frac{1}{\h}\sum_{|\nu'|=0,\zeta'=\pm 1}^{\infty}
|\nu',\zeta',t\rangle \int\limits_0^t d\tau\langle\tau,\zeta',\nu'|
\big[\hat F_1^{(2)} -\hat F_1^{(1)}-{}\\
&-&\displaystyle\frac{\h}{2}(- i\h\pa_\tau + \hat F_0)\big(\frac{1}{2\ve}
\hat Q_1\big)^2-\frac{\h}{2}\big(\frac{1}{2\ve}\hat Q_1\big)^2(- i\h\pa_\tau
+\hat F_0)\big]|\va(\tau)\rangle.
\end{eqnarray*}
Thus the functions (5.5) form the complete orthonormalized set
of solutions of equation (5.3):
$$
\langle\tilde{\cal U}_{\nu'\zeta'}^{(2)}(\vec x,t,\h)|
\tilde{\cal U}_{\nu,\zeta}^{(2)}(\vec x,t,\h)\rangle =\delta_{\nu,\nu'}
\delta_{\zeta,\zeta'}+O(\h^{3/2}),
$$
which can be considered as the relativistic (up to $O(\h^{5/2})$)
generalization of the Pauli equation. In the case $N>2$ the operator
$\hat B^{(N)}$ ($\hat{\tilde\CH}{}^{(N)}$, $\hat{\tilde T}{}^{(N)}$
respectively) can be obtained in the same way as the operator (5.4).

Let us explicitly calculate the operator $\hat\CK_D^{(2)}(t,\h)$
which defines the passage to the semiclassical trajectory-coherent
representation. By (4.6), (5.4), we get:
\begin{eqnarray}
\hat\CK_D^{(2)}(t,\h)&=&\Big\{\Pi_+(t) + \frac{1}{2\ve}\Pi_-(t)\big(\sqrt{\h}
\hat Q_1+ \h\hat Q_2 + \h^{3/2}\hat Q_3 + \h^2\hat Q_4\big)\Big\}\times\cr
&\times&\left(\big(1-\frac{\h}{2}(\frac{1}{2\ve}\hat Q_1)^2\big)\big[
1-i\sqrt{\h}\hat{\tilde\CF}_1- i\h\hat{\tilde\CF}_2 - \h\hat{\tilde\CF}{}_1^2
\big]\right)\hat\CK_D^{(0)}(t,\h),
\end{eqnarray}
where $\hat\CK_D^{(0)}(t,\h)$ is defined in (4.6).
Since the operator $\hat{\tilde T}{}^{(N)}$ is unitary,
the operator (5.6) is also unitary, i.e.:
\begin{eqnarray*}
\lefteqn{\langle(\hat\CK_D^{(2)}(t,\h)|H_{\nu'}\zeta'\rangle)|(\hat\CK_D^{(2)}(t,\h)|
H_\nu\zeta\rangle)\rangle_{L_2}={}}\\
&&=\langle\zeta'H_{\nu'}|H_\nu\zeta\rangle_{L_{\h}^t}+O(\h^{3/2})=
\delta_{\nu,\nu'}\delta_{\zeta,\zeta'} +O(\h^{3/2}).
\end{eqnarray*}

Let $\hat A_t:L_2({\Bbb R}^3,{\Bbb C}^4) \to L_2({\Bbb R}^3,{\Bbb C}^4)$ be a
unitary operator. Then the corresponding operator in
the two-component theory can be always represented in the form:
$$
\hat{\tilde\CA}_+ = \{\hat{\tilde T}{}^{(2)}\}^+ \hat\CA_t
\hat{\tilde T}{}^{(2)} + \hat O(\h^{3/2}) =
\hat a+\langle\vec\sigma,\hat{\vec A}\rangle +\hat O(\h^{3/2}),
$$
where $\hat a$ and $\hat{\vec A}$ are self-adjoint (in $L_2$) operators with
symbols $a(\vec x,\vec p,t)$ and $\vec A(\vec x,\vec p,t)$ respectively.
We find the explicit form of the operator $\vec A_t(\hbar)$ in the
semiclassical TC-representation:
\begin{eqnarray}
\hat\CA'&=&(\hat\CK_D^{(2)}(t,\h))^+\hat\CA_t\hat\CK_D^{(2)}(t,\h) =
(\hat\CK_D^{(0)}(t,\h))^{(+)}(\hat{\tilde \CF}{}^{(2)})^{(+)}
\hat{\tilde\CA}_+\hat{\tilde F}{}^{(2)}\hat\CK_D^{(0)}(t,\h)={}\cr
&=&\Big(1+{\hat D}^1+\frac{1}{2}{\hat D}^2\Big)\big(a(t) +\langle\vec\sigma,
\vec A(t)\rangle\big)- i\sqrt{\h}\Big[{\hat D}^1\big(a(t) +\langle
\vec\sigma,\vec A(t)\rangle\big)\hat\pi_1-{}\cr
&-&\hat\pi_1^+\hat D^1\big(a(t) +\langle\vec\sigma,\vec A(t)\big)\Big] +
\h\langle\vec A(t),\big[\hat\pi_1^+ \vec\sigma\hat\pi_1-{}\cr
&-&i(\vec\sigma\hat\pi_2-\hat\pi_2^+\vec\sigma) -\vec\sigma\hat\pi_1^2 -
(\hat\pi_1^+)^2\vec\sigma\big]\rangle - i\sqrt{\h}\langle\vec A(t),
(\vec\sigma\hat\pi_1-\hat\pi_1^+\vec\sigma)\rangle +\hat O(\h^{3/2}),
\end{eqnarray}
where
$$
\hat\pi_j =\big(\hat\CK_D^{(0)}(t,\h)\big)^{-1}\hat{\tilde\CF}_j
\hat\CK_D^{(0)}(t,\h),\quad j = 1,2,
$$
and $\hat D^j \CA(t)$ are defined (1.14).

In particular, in the semiclassical TC-representation, the operators
of momentum $\hat{\vec p} =-i\h\nabla$, coordinates $\vec x$, and
spin\footnote{In the initial (Schr\"odinger) representation, the
spin operator $\hat{\vec S}=(\h/2)\vec\sigma$ corresponds (up to
$O(\h^{3/2})$) to the three-dimensional unit vector of spin
$(\h/2)\vec\sigma^0=(\h/2)\{\rho_3\vec \Sigma +\rho_1(c/\ve)\hat{\vec\CP}
-c\rho_3\hat{\vec\CP}\langle\vec\Sigma,\hat{\vec\CP}\rangle/[\ve(\ve+m_0 c^2)]\}]$
(see \cite{34, 35}).} $\hat{\vec S}=\h\vec\sigma/2$ have the form:
\begin{eqnarray}
\hat{\vec X}(t,\h)&=&\bigl\{\hat\CK_D^{(0)}(t,\h)\bigr\}^{-1}
\bigl\{\hat{\tilde\CF}{}^{(2)}\bigr\}^{-1}\vec x \hat{\tilde\CF}{}^{(2)}
\hat\CK_D^{(0)}(t,\h)={}\cr
&=&\vec x(t,z_0)+ \Delta\vec{x} - i\sqrt{\h}(\Delta\vec{x}{\hat\pi}_1 -
{\hat\pi}_1^+ \Delta\vec{x}) + \hat O(\h^{3/2}),\\
\hat{\vec P}(t,\h)&=&\bigl\{\hat\CK_D^{(2)}(t,\h)\bigr\}^{-1}
\bigr\{\hat{\tilde\CF}{}^{(2)}\bigr\}^{-1}
\hat{\vec p}\hat{\tilde\CF}{}^{(2)}\hat\CK_t^{(0)}={}\cr
&=&\vec p(t,z_0)+\Delta\hat{\vec p}- i\sqrt{\h}(\Delta\hat{\vec p}
\hat\pi_1-\hat\pi_1^+\Delta\hat{\vec p}) + \hat O(\h^{3/2}),\\
\hat{\vec S}(t,\h)&=&\bigl\{\hat\CK_D^{(2)}(t,\h)\bigr\}^{-1}
\bigl\{\hat{\tilde\CF}{}^{(2)}\bigr\}^{-1}\frac{\h}{2}
\vec\sigma\hat{\tilde\CF}{}^{(2)}\hat\CK_t^{(0)}(\h) =
\frac{\h}{2}\vec\sigma + \hat O(\h^{3/2}).
\end{eqnarray}
Let us write the explicit expression for the Hamiltonian of the
two-component theory $\hat\CH^{(N)}$ (5.2), in which the
operators of the order $\hat O(\h^{3/2})$ are taken into account,
and the expressions for quantum averages of the principal observables
in the theory, namely, for operators of coordinates, momenta and spin
are calculated for one-particle semiclassical TC-states of electron
$|H_\nu,\zeta\rangle$ (4.10):
\begin{eqnarray}
\lefteqn{\hat{\tilde\CH}{}^{(1)} =(1 +\hat\delta^1 + \frac{1}{2!}\hat{\delta}^2 +
\frac{1}{3!}\hat\delta^3)\lambda^{(+)}(t)+{}}\cr
&&{}+\frac{ec\h}{2\ve}\Big\{\Big[(1+\tilde g\gamma)\langle\vec\sigma,
\vec H(t)\rangle +(\frac{1}{\ga}+\tilde g\gamma)\sbet+{}\cr
&&{}+ \frac{\tilde g\gamma}{\ga}\sb\bvht\Big] -\frac{c}{\ve}\Big\langle\vec\sigma,
\big[\vec\beta\frac{\bvet}{\ga}-\vec E(t)\big] \times \big[\vec\beta
\frac{\langle\vec\beta,\hat{\vec\CP}_1\rangle}{\ga}-
\hat{\vec\CP}_1\big]\Big\rangle-{}\cr
&&{}- \Big[\sb\frac{\langle\vec\beta,d\vec H\rangle}{\ga} +{\gamma}^{-1}
\langle\vec\sigma,d\vec H\rangle\Big] -\frac{c}{\ve}\Big[\sb
\frac{\bvht}{\ga} +{\gamma}^{-1}\svht\Big]\langle\vec\beta,
\hat{\vec\CP}_1\rangle+{}\cr
&&{}+\tilde g\gamma\Big[\langle\vec\sigma,\vec\beta\times d\vec E\rangle -
\langle\vec\sigma,d\vec H\rangle + \sb\frac{\langle\vec\beta,d\vec H\rangle}
{\ga}\Big] -\frac{c}{\ve}\tilde g\gamma\bvht\Big[\sb\frac{\langle\hat{\vec\CP}_1,
\vec\beta\rangle}{\ga}-{}\cr
&&{}-\langle\vec\sigma,\hat{\vec\CP}_1\rangle\Big]-\tilde g\gamma\frac{c}{\ve}
\Big\langle\vec\sigma,\Big[\vec\beta\frac{\bvet}{\ga} +
\gamma^{-1}\vec E(t)\Big] \times\Big[\vec\beta\frac{\langle\vec\beta,
\hat{\vec\CP}_1\rangle}{\ga}-\hat{\vec\CP}_1\Big]\Big\rangle\Big\}={}\cr
&&{}=\Big(1+\hat\delta^1+\frac{1}{2!}\hat\delta^2 + \frac{1}{3!}\hat\delta^3\Big)
\lambda^{(+)}(t)+\langle\vec\sigma,\vec D_0(t)\rangle+\langle\vec\sigma,
[D_x(t)\Delta\vec x+ D_p (t) \Delta\hat{\vec p}]\rangle;\\
\lefteqn{D_p (t) \Delta\hat{\vec p}=
-\displaystyle\frac{c}{\ve}\big[\vec\beta\frac{\bvet}{\ga}-\vec E(t)\big]
\times \big[\vec\beta\frac{\langle\vec\beta,\Delta\hat{\vec p}\rangle}{\ga}-
\Delta\hat{\vec p}\big]-{}}\cr
&&{}-\displaystyle\frac{c}{\ve}\Big[\vec\beta\frac{\bvht}{\ga} +{\gamma}^{-1}
\vec H(t)\Big]\langle\vec\beta,\Delta\hat{\vec p}\rangle-{}\cr
&&{}-\displaystyle\frac{c}{\ve}\tilde g\gamma\bvht\Big[\vec\beta\frac{\langle\Delta
\hat{\vec p},\vec\beta\rangle}{\ga}-\Delta\hat{\vec p}\Big]-{}\cr
&&{}-\tilde g\gamma\displaystyle\frac{c}{\ve}\Big[\vec\beta\frac{\bvet}{\ga} +
\gamma^{-1}\vec E(t)\Big] \times\Big[\vec\beta\frac{\langle\vec\beta,
\Delta\hat{\vec p}\rangle}{\ga}-\Delta\hat{\vec p}\Big];\cr
\lefteqn{D_x (t) \Delta\hat{\vec x}=-\displaystyle\frac{c}{\ve}\big[\vec\beta
\frac{\bvet}{\ga}-\vec E(t)\big] \times \big[\vec\beta
\frac{\langle\vec\beta,\Big(-{e\over c}d^1\vec\CA(t)\Big)\rangle}
{\ga}-\Big(-{e\over c}d^1\vec\CA(t)\Big)\big]-{}}\cr
&&{}-\displaystyle\frac{c}{\ve}\Big[\vec\beta\frac{\bvht}{\ga} +{\gamma}^{-1}
\vec H(t)\Big]\langle\vec\beta,\Big(-{e\over c}d^1\vec\CA(t)\Big)\rangle-{}\cr
&&{}-\displaystyle\frac{c}{\ve}\tilde g\gamma\bvht\Big[\vec\beta
\frac{\langle\Big(-{e\over c}d^1\vec\CA(t)\Big),\vec\beta\rangle}{\ga}-
\Big(-{e\over c}d^1\vec\CA(t)\Big)\Big]-{}\cr
&&{}-\tilde g\gamma\displaystyle\frac{c}{\ve}\Big[\vec\beta\frac{\bvet}{\ga} +
\gamma^{-1}\vec E(t)\Big] \times\Big[\vec\beta\frac{\langle\vec\beta,
\Big(-{e\over c}d^1\vec\CA(t)\Big)\rangle}{\ga}-\Big(-{e\over c}d^1
\vec\CA(t)\Big)\Big]-{}\cr
&&{}- \displaystyle\Big[\sb\frac{\langle\vec\beta,d\vec H\rangle}{\ga}
+{\gamma}^{-1}\langle\vec\sigma,d\vec H\rangle\Big] +\tilde g\gamma
\Big[\langle\vec\sigma,\vec\beta\times d\vec E\rangle -\langle\vec\sigma,
d\vec H\rangle + \sb\frac{\langle\vec\beta,d\vec H\rangle}{\ga}\Big]; \cr
\lefteqn{\vec X (t,\zeta,\zeta') =\langle\zeta',H_{\nu} |
\hat{\vec X} (t,\h) | H_{\nu},\zeta\rangle={}} \cr
&&{}=\vec x(t,z_0)\delta_{\zeta,\zeta'} -i\sqrt{\h}
\langle\zeta',H_{\nu} |(\Delta\vec x\hat\pi_1 -
\hat\pi_1^+ \Delta\vec x) | H_\nu,\zeta\rangle +\hat O(\h^2);\\
\lefteqn{\vec P(t,\zeta,\zeta') = \langle\zeta',H_\nu|\hat{\vec P}(t,\h)|
H_\nu,\zeta\rangle={}}\cr
&&{}= \vec p(t,z_0)\delta_{\zeta,\zeta'} -i\sqrt{\h}\langle\zeta',H_\nu|
(\Delta\vec p{}'\hat\pi_1 - \hat\pi_1^+\Delta\vec p{}')| H_\nu,\zeta\rangle
+ \hat O(\h^2);\\
\lefteqn{\vec S(t,\zeta,\zeta') = \langle\zeta',H_{\nu}|\hat{\vec S}(t,\h)|
H_\nu,\zeta\rangle =\frac{\h}{2}\vec\eta(t,\zeta,\zeta').}
\end{eqnarray}
Here $\Delta\hat{\vec p}{}'=-i\h\nabla+Q(t)\Delta\vec x$, $\tilde g=
\frac{1}{2}(g-2)$,   $\vec\eta(t,\zeta,\zeta')={\cal U}^+(t,\zeta')\vec\sigma
{\cal U}(t,\zeta)$ is the solution of the Bargmann--Michel--Telegdi equation
\cite{20} in the rest system
$$
\dot{\vec\eta} = \frac{2}{\h}\vec\eta\times\vec{\cal D}_0(t,z_0),
$$
where the vector $\vec{\cal D}_0(t,z_0)$ is defined in (4.8).

By using the function $|H_\nu,\zeta\rangle$ (4.10), it is easy to
calculate (with the same precision $O(\h^2)$, $\h\to0$) the
correlation matrix \cite{15}, which describes quantum fluctuations of
dynamical variables $\hat x_j(t,\h)$, $\hat p_j(t,\h)$ with
respect to their average values (5.12), (5.14), $\sigma_{x_i,x_j}$,
$\sigma_{p_i,p_j}$, and their correlation $\sigma_{p_i,x_j}$, $i,j=1,2,3$.
Here
$$
\sigma_{AB} = \frac{1}{2}\langle(\hat A\hat B + \hat B\hat A)\rangle -
\langle\hat A\rangle\langle\hat B\rangle.
$$
We have
\begin{eqnarray}
&&\sigma_{xx} =\frac{\h}{4}\big[C(t)D_\nu^{-1}C^+(t)+C^*(t)D_\nu^{-1}
C^t(t)\big],\cr
&&\sigma_{pp} =\frac{\h}{4}\big[B(t)D_\nu^{-1}B^+(t)+
B^*(t)D_\nu^{-1}B^t(t)\big],\\
&&\sigma_{px} =\frac{\h}{4}\big[B(t)D_\nu^{-1}C^+(t)+B^*(t)D_\nu^{-1}
C^t(t)\big],\nonumber
\end{eqnarray}
where $C(t)$ and $B(t)$ are defined in (I.1.3), and $D_\nu^{-1}$
denotes the matrix
$$
D_\nu^{-1} = \left\|\frac{2\nu_k+1}{\Im\,b_k}\delta_{jk}\right\|.
$$

Differentiating the relations (5.12)--(5.15) with respect to $t$ and
writing the right-hand sides of obtained relations in terms of
variables $\dot{\vec X}$, $\vec P$, $\vec\eta$, $\Delta_2$, we get
(up to $O(\h^{3/2})$)
\begin{eqnarray}
&&\dot z = J\pa_z\lambda^{(+)}(z,t)+\frac{1}{2}
\langle\pa_z,\Delta_2\pa_z\rangle J\pa_z\lambda^{(+)}(z,t)+JD_z(z,t)\vec\eta,\cr
&&\dot\Delta_2 = J(\lambda^{(+)})''_{zz}(z,t)\Delta_2-\Delta_2
(\lambda^{(+)})''_{zz}(z,t)J, \qquad \Delta_2^t=\Delta_2 ,\\
&&\dot{\vec\eta}=\displaystyle\frac{2}{\h}\vec\eta\times\vec{\cal D}_0(t,z_0),\nonumber
\end{eqnarray}
where
$$
z=(\vec P(t,\h) ,\vec X(t,\h)) , \qquad
\Delta_2 = \left(\begin{array}{cc}\sigma_{pp}&\sigma_{px}\\
\sigma_{xp}&\sigma_{xx}\end{array}\right),
$$
$$
D_z\vec\eta = (D_p\vec\eta,D_x\vec\eta), \qquad \vec \beta =\frac 1 c
\lambda_{\vec p} (z,t),
$$
vector $\vec D_0$ is defined in (4.8), and $3\times 3$ matrices $D_p$, $D_x$ - in
(5.11), $\vec P(t,\hbar)$, $\vec X(t,\hbar)$ in (5.12), (5.13).
The system (5.16) is a closed system of ordinary differential equations
for quantum averages in Dirac theory. The problem of correspondence
between the obtained equations and the well-known equations for a particle
with spin (e.g., Frenkel etc. \cite{36, 37}) requires a more detail consideration.
The initial conditions for system (5.16) is choosen as follows:
\begin{equation}
\begin{array}{c} z\big|_{t=0}=z_0, \qquad \Delta_2\big|_{t=0}=\Delta_2^0,\\
\vec\eta\big|_{t=0}=\displaystyle\frac{1+\zeta\zeta'}{2}\zeta\vec\ell+
\frac{1-\zeta\zeta'}{2}\frac{\vec\ell\times(\vec k\times\vec\ell)+
i\zeta\vec\ell\times\vec k}{\sqrt{1-(\vec\ell,\vec k)^2}},\end{array}
\end{equation}
where $\vec k=(0,0,1)$, $\vec\ell$ and $\Delta_0^2$ are defined in (4.4)
and (I.5.23), respectively. Let us write the system (5.16) in the form
\begin{eqnarray}
&&\dot z = J\pa_z\lambda^{(+)}(z,t) + \displaystyle\frac{1}{2} \lan\pa_z,
A\Delta^0_2A^+\pa_z\ran J\pa_z\lambda^{(+)}(z,t)+JD_z(z,t)
{\cal U}^+\vec\sigma{\cal U},\cr
&&\dot A= J (\lambda^{(+)})''_{zz}(z,t) A,\\
&&\Big[i\displaystyle\frac{d}{dt}+{1\over\h}\lan\vec\sigma,\vec{\cal D}_0\ran
\Big]u=0.\nonumber
\end{eqnarray}
The initial conditions for spinor $\cal U$ are defined in (4.4) and
\[ A\big|_{t=0}=\left(\begin{array}{cc}B_0 & B_0^*\\
C_0 & C_0^*\end{array}\right) \]
where $3\times3$ matrices $B_0$ and $C_0$ satisfy the conditions (I.1.4)
and (I.1.5).

\section{Conclusions}

Let us briefly review our consideration. To our opinion, from
the physical point of view, the ``passage to the classics'' in
quantum mechanics  must inevitably cause the introduction of the
notion of classical trajectory, which primarily (on the
postulational level) is alien to quantum mechanics and must be
taken from the outside. It is essential that one can construct
the complete set of approximate solutions of the Dirac equation with the
property: as $\h\to 0$, the quantum-mechanical average values
of coordinates and momenta are general solutions of classical
Hamiltonian equations. In literature \cite{38} it was repeated
over and over  that this possibility is not obvious. Usually (see, for
example, \cite{34}, p. 69--70) the authors restrict themselves to
a verbal formulation of conditions imposed on the relativistic wave
function of semiclassical type, having implicitly in mind that these
conditions can be met without any essential difficulties. However,
as a rule, such states were not presented. Our consideration shows
that in order to construct such states explicitly, one can  use the
method of complex germ \cite{4}.

Not only the principal possibility of obtaining approximate (in
$\h\to0$) solutions of  the  Dirac equations (up to $O(\h^{N/2})$)
for any $N$ is shown, but also a constructive method for obtaining
the corresponding higher approximations is given. It is essential
that the expansion into an asymptotic series with respect to $\h$
contains half-integer powers of $\h$, i.e., there are series in
$\sqrt{\h}$ (in contrast to the standard semiclassical expansion
in $\h$, $\h\to0$, given in all manuals of quantum mechanics).

The possibility to construct such states (called trajectory-coherent
states) explicitly leads to nontrivial conclusions. For example, one
can obtain, in the most transparent and natural way, the ``classical
equations of motion'' for average values of quantities, which cannot
be exactly well-defined in the classical sense (e.g., spin). It seems
natural that, for the classical vector of spin, we obtain the
Bargman--Michel--Telegdi equation. However,  in the case of arbitrary (and
not only homogeneous) electromagnetic fields, one can explicitly
show, which is nontrivial, that in this equations the fields must be
taken on classical trajectories. Up to now (see, for example, \cite{34}),
this fact was justified only by verbal arguments.

In conclusion, we note that, by using the Maslov complex
canonical operator, one can go over to the approximate
trajectory-coherent representation, in which the classical
trajectory is considered already for the approximate Hamiltonian
of quantum theory.

\section*{Acknowledgements}

This work was supported by the International Science Foundation, under Grants
No. p98-138 and No. p98-455, and Russian Fundamental Research Foundation,
under Grants No. 97-02-16279 and No. 98-02-16195. We thank
M.A. Shishkova and M.F. Kondrat'eva for assistance in preparation of this
manuscript and useful discussions.

\section*{Appendix A. Properties of matrices $\Pi_\pm(t)$
and $\lambda^{(+)}_{pp}(t)$}
\def\thesection{A}

\begin{Prop} The following  relations hold{\rm:}
\begin{eqnarray}
&&[\lambda^{(+)}_{pp}(t)]^{1/2}=\frac{c}{\sqrt{\ve(t)}}\left\|
\frac{\beta_i\beta_j}{\ga}-\delta_{ij}\right\|_{3\times3},\\
&&[\lambda^{(+)}_{pp}(t)]^{-1/2}=-\frac{\sqrt{\ve(t)}}{c}\left\|
\delta_{ij}+\frac{\gamma\beta_i\beta_j}{\ga}\right\|_{3\times3},\\
&&[\lambda^{(+)}_{pp}(t)]^{-1}=\frac{\ve(t)}{c^2}\|\delta_{jk}+
\gamma^2\beta_j\beta_k\|_{3\times3},
\end{eqnarray}
where
$$
\vec\beta=\frac{1}{c}\dot{\vec x}(t,z_0), \qquad
\gamma^{-1}=(1-\vec\beta^2)^{1/2}, \qquad
\lambda^{(+)}_{pp}=\frac{c^2}{\ve}\|\delta_{ij}-\beta_i\beta_j\|.
$$
\end{Prop}
{\bf Proof}. We can verify directly that
$$
\sum_{i=1}^3\frac{c}{\sqrt{\ve(t)}}\left(\frac{\beta_i\beta_j}
{\ga}-\delta_{ij}\right)\frac{c}{\sqrt{\ve(t)}}\left(
\frac{\beta_i\beta_k}{\ga}-\delta_{ik}\right)=
\leqno{{\rm i)}}$$
$$
=\frac{c^2}{\ve(t)}\left(\frac{\beta_j\beta_k}{\ga}
\left(\frac{\vec\beta^2}{\ga}-2\right)+\delta_{jk}\right)=
$$
$$
=\frac{c^2}{\ve(t)}(\delta_{jk}-\beta_j\beta_k)=\lambda^{(+)}_{p_jp_k}(t),
\qquad \beta^2=1-\gamma^{-2}.
$$
$$
\sum_{i=1}^3\frac{c}{\sqrt{\ve(t)}}\left(\frac{\beta_j\beta_i}
{\ga}-\delta_{ij}\right)\left(-\frac{\sqrt{\ve(t)}}{c}\right)
\left(\delta_{ik}+\frac{\gamma\beta_i\beta_k}{\ga}\right)=
\leqno{\rm ii)}$$
$$
=-\left[\frac{\beta_j\beta_k}{\ga}\left(\frac{\gamma\vec\beta^2}
{\ga}-\gamma+1\right)-\delta_{jk}\right]=\delta_{jk}.
$$
$$
\sum_{i=1}^3\frac{c^2}{\ve(t)}(\delta_{ij}-\beta_i\beta_j)
\frac{\ve(t)}{c^2}(\delta_{ik}+\gamma^2\beta_i\beta_k)=
\delta_{jk}+\beta_j\beta_k(\gamma^2-1-\gamma^2\vec\beta^2)=\delta_{jk},
\leqno{\rm iii)}$$
as was to be proved.
\begin{Prop} The following relation holds{\rm :}
\begin{equation}
\CH_0(t)\Pi_\pm(t)=\lambda^{(\pm)}(t)\Pi_\pm(t),
\end{equation}
where  $\CH_0(t)$, $\lambda^{(\pm)}(t)$, and $\Pi_\pm(t)$
are defined in $(2.2)$, $(2.7)$, and $(2.8)$, respectively.
\end{Prop}
{\bf Proof}. Since $\dot{\vec\beta}=c\vec\CP/\ve$, we get
\begin{eqnarray*}
\CH_0(t)\Pi_+(t)&=&\frac{1}{\ve\sqrt{2+2\gamma^{-1}}}\left(\begin{array}{cc}
e\Phi+m_0c^2 & c\sp\\
c\sp & e\Phi-m_0c^2 \end{array}\right)\left(\begin{array}{c}
\ve+m_0c^2 \\ c\sp\end{array}\right)={}\\
&=&\frac{1}{\ve\sqrt{2+2\gamma^{-1}}}\left(\begin{array}{c}
(e\Phi+m_0c^2)(\ve+m_0c^2)+c^2\sp^2\\
c\sp(\ve+m_0c^2+e\Phi-m_0c^2)\end{array}\right)=(\ve+e\Phi)\Pi_+(t).
\end{eqnarray*}
Here we took into account that $c^2\sp^2=c^2\vec\CP^2=\ve^2-m_0^2c^4$.
For the lower index, relation (A.4) can be proved similarly.
\begin{Prop} The following relation holds{\rm :}
\begin{eqnarray}
\rho_1\langle\vec\Sigma,\vec\CP\rangle\Pi_\pm(t)&=&\langle\vec\alpha,
\vec\CP\rangle\Pi_\pm(t)=\pm\bp\Pi_\pm(t)+\Pi_\mp(t)\left(\sb\frac{\bp}{\ga}-
\sp\right)={}\cr
&=&\pm\bp\Pi_\pm(t)+\frac{\sqrt{\ve(t)}}{c}\Pi_\mp(t)\langle\vec\sigma,
(\lambda^{(+)}_{pp}(t))^{1/2}\vec\CP\rangle.
\end{eqnarray}
\end{Prop}
{\bf Proof}. Actually,
\begin{eqnarray*}
\lefteqn{\langle\vec\alpha,\vec\CP\rangle\Pi_+(t)=\frac{1}{\sqrt{2+2\gamma^{-1}}}
\left(\begin{array}{c}\sp\sb\\ \sp(\ga)\end{array}\right).}\\
\lefteqn{\sp\sb=\bp+ i\langle\vec\sigma,\vec\CP\times\vec\beta\rangle
=[1+(\ga)-(\ga)]\bp-{}}\\
&&{}-i\langle\vec\sigma,\vec\CP\times\vec\beta\rangle=
(\ga)\bp+[(1-\gamma^{-1})\bp-\sb\sp]={}\\
&&{}=\left|(1-\gamma^{-1})=\frac{\vec\beta^2}{\ga}=
\frac{1}{\ga}\sb^2\right|={}\\
&&{}=(\ga)[\bp]+\sb\left[\sb\frac{\bp}{\ga}-\sp\right].\\
\lefteqn{(\ga)\sp=(\ga)\sp-\bp\sb+\bp\sb={}}\\
&&{}=\sb[\bp]-(\ga)\left[\sb\displaystyle\frac{\bp}{\ga}-\sp\right].
\end{eqnarray*}
Thus, we obtain the first relation in (A.5) for upper indices.
Similarly
\begin{eqnarray*}
\lefteqn{\langle\vec\alpha,\vec\CP\rangle\Pi_-=\displaystyle\frac{1}
{\sqrt{2(\ga)}}\left(\begin{array}{c}-\sp(\ga)\\\sp\sb\end{array}\right);}\\
\lefteqn{-\sp(\ga)=\langle\vec\beta,\vec\CP\rangle\langle\vec\beta,
\vec\sigma\rangle+\bp\sb-(\ga)\sp={}}\\[6pt]
&&{}=\sb[-\bp]+(\ga)\left[\sb\displaystyle\frac{\bp}{\ga}-\sp\right];\\
\lefteqn{\sp\sb=(\ga)\bp+(\ga)\bp-\sb\sp={}}\\[6pt]
&&{}=(-1-\gamma^{-1})[-\bp]+\sb\left[\sb\displaystyle\frac{\bp}{\ga}-\sp\right],
\end{eqnarray*}
as was to be proved. The second relation in (A.5) immediately follows
from (A.1).
\begin{rem} If in the matrices $\Pi_\pm(t)=\Pi_\pm(\vec p,\vec x,t)
\bigg|_{\vec p=\vec p(t),\vec x=\vec x(t)}$ trajectory $\vec x(t)$,
$\vec p(t)$ is not classical, then in the obtained expressions we must
put $\vec\beta=c\vec\CP/\ve$.
\end{rem}
\begin{Prop} The following relation holds{\rm :}
\begin{eqnarray}
\dot\Pi_\pm(t)&=&\Pi_\pm(t)\left[\frac{i}{2}\frac
{\sbb}{\ga}\right]\mp\frac{1}{2}\Pi_\mp(t)
\left[\sb\frac{\gamma\bdb}{\ga}+\sdb\right]={}\cr
&=&\frac{i}{2(\ga)}\Pi_\pm(t)\sbb\pm\frac{c}{2\sqrt{\ve(t)}}
\Pi_\pm(t)\langle\vec\sigma,(\lambda^{(+)}_{pp}(t))^{-1/2}
\dot{\vec\beta}\rangle.
\end{eqnarray}
\end{Prop}
{\bf Proof}. The latter relation in Eq. (A.6) is a direct consequence
of Eq. (A.2). We consider
$$
\frac{d}{dt}\Pi_+(t)=\frac{1}{\sqrt{2}}\left(\begin{array}{c}
\displaystyle\frac{d}{dt}\sqrt{\ga}\\[6pt]
\displaystyle\frac{d}{dt}\displaystyle\frac{1}{\sqrt{\ga}}\sb\end{array}
\right);
\leqno{{\rm i)}}$$
\begin{eqnarray*}
\frac{d}{dt}\sqrt{\ga}&=&\frac{1}{2}\frac{d\gamma^{-1}/dt}{\sqrt{\ga}}=
\frac{1}{2}\displaystyle\frac{1}{\sqrt{\ga}}\frac{d\gamma^{-1}}{dt}\left(
\frac{\vec\beta^2}{\ga}+\gamma^{-1}\right)=\\
&=&\frac{1}{2\sqrt{\ga}}\left[i\sbb+\frac{\vec\beta^2}{\ga}
\frac{d\gamma^{-1}}{dt}+\gamma^{-1}\frac{d\gamma^{-1}}{dt}-i\sbb\right]=\\
&=&\left|\frac{d\gamma^{-1}}{dt}=-\gamma\bdb\right|=\frac{1}{\sqrt{\ga}}\left\{
(\ga)\left[\frac{i}{2}\frac{\sbb}{(\ga)}\right]-\right.\\
&-&\left.\frac{1}{2}\sb\left[\sb\displaystyle\frac{\gamma\bdb}
{\ga}-\sdb\right]\right\};\\
\frac{d}{dt}\frac{\sb}{\sqrt{\ga}}&=&\frac{1}{2\sqrt{\ga}}\left\{2\sdb-
\frac{1}{\ga}\sb\frac{d\gamma^{-1}}{dt}\right\}=\\
&=&\frac{1}{2\sqrt{\ga}}\left\{\sdb\left(\frac{\vec\beta^2}{\ga}+(\ga)\right)+
\sb\left(\frac{\gamma^{-1}}{\ga}-1\right)\right\}=\\
&=&\big|\vec\beta^2\sdb-\sb\bdb=-\langle\vec\sigma,\vec\beta\times(\vec\beta
\times\dot{\vec\beta})\rangle=i\sb\sbb\big|=\\
&=&\frac{1}{2\sqrt{\ga}}\left\{i\frac{\sb\sbb}{\ga}+(\ga)\sdb-\sb
\frac{d\gamma^{-1}}{dt}\right\}=\\
&=&\frac{1}{\sqrt{\ga}}\left\{\sb\left[\frac{i}{2}\frac{\sbb}{\ga}\right]+
\frac{1}{2}(\ga)\left[\sb\frac{\gamma\bdb}{\ga}+\sdb\right]\right\}.
\end{eqnarray*}
Similarly,
$$
\frac{d}{dt}\Pi_-(t)=\frac{1}{\sqrt{2}}\left(\begin{array}{c}
\displaystyle\frac{d}{dt}\displaystyle\frac{\sb}{\sqrt{\ga}}\\[6pt]
-\displaystyle\frac{d}{dt}\sqrt{\ga}\end{array}\right);
\leqno{{\rm ii)}}$$
\begin{eqnarray*}
\frac{d}{dt}\,\frac{\sb}{\sqrt{\ga}}&=&\frac{1}{2\sqrt{\ga}}\left\{\sb\left[i
\frac{\sbb}{\ga}\right]+\right.\\
&&\left.+(\ga)\left[\sb\frac{\gamma\bdb}{\ga}+\sdb\right]\right\};\\
-\frac{d}{dt}\sqrt{\ga}&=&\frac{1}{2\sqrt{\ga}}\left\{-(\ga)\left[
i\frac{\sbb}{\ga}\right]+\right.\\
&&\left.+\sb\left[\sb\frac{\gamma\bdb}{\ga}-\sdb\right]\right\},
\end{eqnarray*}
as was to be proved.
\begin{Prop} The following relation holds{\rm :}
\begin{eqnarray}
\rho_3\langle\vec\Sigma,\vec H\rangle\Pi_\pm(t)&=&\mp\Pi_\pm(t)\left[\sb
\frac{\langle\vec\beta,\vec H\rangle}{\ga}-\langle\vec\sigma,\vec H\rangle
\right]+\Pi_\mp(t)\langle\vec\beta,\vec H\rangle={}\cr
&=&\mp\frac{\sqrt{\ve(t)}}{c}\Pi_\pm(t)\langle\vec\sigma,(\lambda^{(+)}_{pp}
(t))^{-1/2}\vec H\rangle+\Pi_\mp(t)\langle\vec\beta,\vec H\rangle.
\end{eqnarray}
\end{Prop}
{\bf Proof}. The latter relation in (A.7) follows directly from Eq.
(A.1). Precisely as in Property 3, we get
$$
\rho_3\langle\vec\Sigma,\vec H\rangle\Pi_+(t)=\frac{1}{\sqrt{2(\ga)}}
\left(\begin{array}{c}\langle\vec\sigma,\vec H\rangle(\ga)\\
-\langle\vec\sigma,\vec H\rangle\sb\end{array}\right);
\leqno{{\rm i)}}$$
$$
\langle\vec\sigma,\vec H\rangle(\ga)=-(\ga)\left[\sb\frac{\langle
\vec\beta,\vec H\rangle}{\ga}-\langle\vec\sigma,\vec H\rangle\right]+
\sb\langle\vec\beta,\vec H\rangle;
$$
$$
-\langle\vec\sigma,\vec H\rangle\sb=-\sb\left[\sb\frac{\langle
\vec\beta,\vec H\rangle}{\ga}-\langle\vec\sigma,\vec H\rangle\right]-
(\ga)\langle\vec\beta,\vec H\rangle;
$$
$$
\rho_3\langle\vec\Sigma,\vec H\rangle\Pi_-(t)=\frac{1}{\sqrt{2(\ga)}}
\left(\begin{array}{c}\langle\vec\sigma,\vec H\rangle\sb\\
(\ga)\langle\vec\sigma,\vec H\rangle\end{array}\right).
\leqno{{\rm ii)}}$$
Further, the proof coincides with that of Property 3.

\begin{Prop} The following relation holds{\rm :}
\begin{eqnarray}
\rho_2\langle\vec\Sigma,\vec E\rangle\Pi_\pm(t)&=&\Pi_\pm(t)(-\langle
\vec\sigma,\vec\beta\times\vec E\rangle)\mp i\Pi_\mp(t)\left[\sb
\frac{\langle\vec\beta,\vec E\rangle}{\ga}+\gamma^{-1}\langle
\vec\sigma,\vec E\rangle\right]={}\cr
&=&-\Pi_\pm(t)\sbe\pm i\gamma^{-1}\frac{c}{\sqrt{\ve(t)}}\Pi_\mp(t)
\langle\vec\sigma,(\lambda^{(+)}_{pp}(t))^{-1/2}\vec E\rangle.
\end{eqnarray}
\end{Prop}
{\bf Proof}. The latter equality in (A.8) follows from (A.2).
We consider the relations
$$
\rho_2\langle\vec\Sigma,\vec E\rangle\Pi_+(t)=\frac{i}{\sqrt{2(\ga)}}
\left(\begin{array}{c}-\langle\vec\sigma,\vec E\rangle\sb\\
\langle\vec\sigma,\vec E\rangle(\ga)\end{array}\right);
\leqno{{\rm i)}}$$
$$
\rho_2\langle\vec\Sigma,\vec E\rangle\Pi_-(t)=\frac{i}{\sqrt{2(\ga)}}
\left(\begin{array}{c}-\langle\vec\sigma,\vec E\rangle(\ga)\\
\langle\vec\sigma,\vec E\rangle\sb\end{array}\right);
\leqno{{\rm ii)}}$$
\begin{eqnarray*}
\lefteqn{-i\langle\vec\sigma,\vec E\rangle\sb=-i\langle\vec\beta,
\vec E\rangle+\langle\vec\sigma,\vec E\times\vec\beta\rangle=}\\
&&=-i(\ga)\langle\vec\beta,\vec E\rangle+i\gamma^{-1}\big(\langle\vec\beta,
\vec E\rangle+i\sbe\big)+(\ga)\langle\vec\sigma,\vec E\times\vec\beta\rangle=\\[6pt]
&&=(\ga)[-\sbe]-i\sb\left[\displaystyle\frac{\langle\vec\beta,\vec E\rangle}{\ga}+\gamma^{-1}
\langle\vec\sigma,\vec E\rangle\right];\\[6pt]
\lefteqn{i\langle\vec\sigma,\vec E\rangle(\ga)=i\left\{\displaystyle\frac{\ga}
{\gamma}+\vec\beta^2\right\}\langle\vec\sigma,\vec E\rangle=}\\[6pt]
&&=-i(\sb\langle\vec\sigma,\vec E\rangle-\langle\vec\sigma,
\vec E\rangle\vec\beta^2)+i\sb\langle\vec\sigma,\vec E\rangle+
\displaystyle\frac{\ga}{\gamma}\langle\vec\sigma,\vec E\rangle=\\[6pt]
&&=\sb[-\sbe]+i(\ga)\left[\sb\displaystyle\frac{\langle\vec\sigma,
\vec E\rangle}{\ga}+\gamma^{-1}\langle\vec\sigma,\vec E\rangle\right],
\end{eqnarray*}
as was to be proved.

\begin{Prop} The following relation holds{\rm :}
\begin{eqnarray}
\langle\vec\Sigma,\vec S\rangle\Pi_\pm(t)&=&\Pi_\pm(t)\left[\gamma^{-1}
\langle\vec\sigma,\vec S\rangle+\frac{\langle\vec\beta,\vec S\rangle}{\ga}
\sb\right]\pm\Pi_\mp(t)[i\langle\vec\sigma,\vec\beta\times\vec S\rangle]={}\cr
&=&-\frac{c\gamma^{-1}}{\sqrt{\ve(t)}}\Pi_\pm(t)\langle\vec\sigma,
(\lambda^{(+)}_{pp}(t))^{-1/2}\vec S\rangle\pm i\Pi_\mp(t)\langle\vec\sigma,
\vec\beta\times\vec S\rangle.
\end{eqnarray}
\end{Prop}
{\bf Proof}. Since
$$
\langle\vec\Sigma,\vec S\rangle\Pi_+(t)=\frac{1}{\sqrt{2(\ga)}}
\left(\begin{array}{c}\langle\vec\sigma,\vec S\rangle(\ga)\\
\langle\vec\sigma,\vec S\rangle\sb\end{array}\right);
$$
$$
\langle\vec\Sigma,\vec S\rangle\Pi_-(t)=\frac{1}{\sqrt{2(\ga)}}
\left(\begin{array}{c}-\langle\vec\sigma,\vec S\rangle\sb\\
\langle\vec\sigma,\vec S\rangle(\ga)\end{array}\right),
$$
the further proof coincides with that of Property 6.

\begin{Prop} The following relation holds{\rm :}
\begin{equation}
\rho_3\Pi_\pm(t)=\pm\gamma^{-1}\Pi_\pm(t)+\Pi_\mp(t)\sb.
\end{equation}
\end{Prop}
{\bf Proof}.
$$
\rho_3\Pi_+(t)=\frac{1}{\sqrt{2(\ga)}}\left(\begin{array}{c}\ga\\
-\sb\end{array}\right);
\leqno{{\rm i)}}$$
$$
(\ga)=\gamma^{-1}(\ga)+\beta^2=(\ga)\gamma^{-1}+(\sb)^2;
$$
$$
-\sb=\gamma^{-1}\sb-(\ga)\sb,
$$
as was to be proved.
$$
\rho_3\Pi_-(t)=\frac{1}{\sqrt{2(\ga)}}\left(\begin{array}{c}\sb\\
\ga\end{array}\right).
\leqno{{\rm ii)}}$$
{\bf Proof} is similar to the preceding one.
\begin{Prop} The following relation holds{\rm :}
\begin{equation}
\rho_2\Pi_\pm(t)=\mp i\Pi_\mp(t).
\end{equation}
\end{Prop}
{\bf Proof} directly follows from the definition of matrices
$\Pi_\pm(t)$.

\begin{Prop} The following relation holds{\rm :}
\begin{equation}
\rho_1\Pi_\pm(t)=\pm\Pi_\pm(t)\sb-\gamma^{-1}\Pi_\mp(t).
\end{equation}
\end{Prop}
{\bf Proof}. Multiplying the left- and right-hand sides of (A.10)
by $i\rho_2$ and transforming the right-hand side of the
obtained expression according to (A.11), we get  (A.12).
\begin{Prop} For matrices $\Pi_\pm(t)$, the orthonormality and
completeness relations hold{\rm:}
\begin{equation}
\Pi^t_\pm(t)\Pi_\pm(t)={\bf I}_{2\times2}, \qquad
\Pi^t_\pm(t)\Pi_\mp(t)=0_{2\times2};
\end{equation}
\begin{equation}
\Pi_+(t)\Pi^t_+(t)+\Pi_-(t)\Pi^t_-(t)={\bf I}_{4\times4}.
\end{equation}
\end{Prop}
{\bf Proof}. By the straightforward verification, we get
\begin{eqnarray*}
\Pi^t_\pm(t)\Pi_\pm(t)&=&\frac{1}{2(\ga)}[(\ga)^2+(\sb)^2]={}\\
&=&\frac{1}{2(\ga)}(1+2\gamma^{-1}+\gamma^{-2}+\vec\beta^2)
{\bf I}_{2\times2}={\bf I}_{2\times2};\\
\Pi^t_\pm(t)\Pi_\mp(t)&=&\frac{\pm1}{2(\ga)}[\sb(\ga)-(\ga)\sb]=0;\\
\Pi_+(t)\Pi^t_+(t)+\Pi_-(t)\Pi^t_-(t)&=&\frac{1}{2(\ga)}\left\{\left(
\begin{array}{cc}(\ga)^2 & \sb(\ga)\\
\sb(\ga) & \vec\beta^2\end{array}\right)\right.+{}\\
&+&\left.\left(\begin{array}{cc}\vec\beta^2 & -\sb(\ga)\\
-\sb(\ga) & (\ga)^2\end{array}\right)\right\}={\bf I}_{4\times4},
\end{eqnarray*}
as was to be proved.

\section*{Appendix B. The Heisenberg equations for the
polarization operator \boldmath$\hat S_\mu$}
\setcounter{Prop}{0}
\setcounter{equation}{0}
\def\thesection{B}

Let us write the Heisenberg equation for the polarization operator:
\begin{equation}
{d\over dt}\langle\Psi|\hat S_\mu|\Psi\rangle_D=
\langle\Psi|{\pa\hat S_\mu\over\pa t}|\Psi\rangle_D+{ i\over\h}
\langle\Psi|[\hat\CH_D,\hat S_\mu]|\Psi\rangle_D,
\end{equation}
where
\begin{equation}
\hat S_\mu=(\hat S_0,\hat{\vec S}),\qquad \hat S_0={1\over m_0c}
\langle\vec\Sigma,\hat{\vec\CP}\rangle,\qquad \hat{\vec S}=\rho_3\vec\Sigma+
{1\over m_0c}\rho_1\hat{\vec\CP},
\end{equation}
\begin{equation}
\begin{array}{c}\hat\CH_D=\hat\CH_0- i\h\hat\CH_1,\qquad
\hat\CH_0=c\langle\vec\alpha,\hat{\vec\CP}\rangle+\rho_3m_0c^2+e\Phi,\\[6pt]
\hat\CH_1=\displaystyle\frac{ ie_0(g-2)}{4m_0c}[\rho_3\langle
\vec\Sigma,\vec H\rangle+\rho_2\langle\vec\Sigma,\vec E\rangle].\end{array}
\end{equation}
For this purpose we need following auxillary propositions:
\begin{Lem} The following relation holds{\rm :}
\begin{equation}
{\pa\hat S_0\over\pa t}+{ i\over\h}[\hat\CH_0,\hat S_0]_-=
{e\over m_0c}\langle\vec\Sigma,\vec E\rangle.
\end{equation}
\end{Lem}
{\bf Proof}. Actually,
$$
{\pa\hat S_0\over\pa t}=-{e\over m_0c^2}\langle\vec\Sigma,{\pa\vec\CA
\over\pa t}\rangle.
$$
$$
[\hat\CH_0,\hat S_0]_-={ ie\h\over m_0c}\langle\vec\Sigma,
\nabla\Phi\rangle.
$$
Since $\vec E=-\nabla\Phi-{1\over c}{\pa\vec\CA\over\pa t}$
the property is proved.
\begin{Lem} The following relation holds{\rm :}
\begin{equation}
{\pa\hat{\vec S}\over\pa t}+{ i\over\h}[\hat\CH_0,\hat{\vec S}]_-=
{e\over m_0c}(\rho_1\vec E+\vec H\times\vec\Sigma).
\end{equation}
\end{Lem}
{\bf Proof}. Actually,
$$
{\pa\hat{\vec S}\over\pa t}=-{e\over m_0c^2}\rho_1{\pa\vec\CA\over\pa t},
$$
\begin{eqnarray*}
[\hat\CH_0,\rho_3\vec\Sigma]_-&=&c\{(i\rho_2)(\hat{\vec\CP}-
 i\vec\Sigma\times\hat{\vec\CP}- i\rho_2(\hat{\vec\CP}+
 i\vec\Sigma\times\hat{\vec\CP})\}=\\
&=&-2 ic\rho_2\hat{\vec\CP},
\end{eqnarray*}
$$
{1\over m_0c}[\hat\CH_0,\rho_1\hat{\vec\CP}]=2 ic\rho_2\hat{\vec\CP}+
{i\h e\over m_0c}\rho_1\nabla\Phi+{1\over m_0}[\langle\vec\Sigma,
\hat{\vec\CP}\rangle,\hat{\vec\CP}]_-.
$$
Since $[\hat\CP_k,\hat\CP_l]=(i\h e/c)(\CA_{l,k}-\CA_{k,l})=
(i\h e/c)\ve_{lkj}H_j$, where $\ve_{lkj}$ is absolutely antisymmetric tensor,
the property is proved.
\begin{Lem} The following relation holds{\rm :}
\begin{eqnarray}
[\hat\CH_1,\hat S_0]_-&=&-{e_0(g-2)\over 2(m_0c)^2}\biggl\{\rho_3\langle
\vec\Sigma,\vec H\times\hat{\vec\CP}\rangle+\rho_2\langle\vec\Sigma,
\vec E\times\hat{\vec\CP}\rangle+{}\cr
&+&{\h\over 2}(\rho_3\div\vec H+\rho_2\div\vec E+ i\rho_3\langle\vec\Sigma,
\rot\vec H\rangle+ i\rho_3\langle\vec\Sigma,\rot\vec E\rangle)\biggr\}.
\end{eqnarray}
\end{Lem}
{\bf Proof}. Actually,
$$
[\CH_1,\hat S_0]_-={ ie_0(g-2)\over 2m_0c}{1\over m_0c}\biggl\{
\rho_3[\langle\vec\Sigma,\vec H\rangle\langle\vec\Sigma,\hat{\vec\CP}\rangle]_-
+\rho_2\langle\vec\Sigma,\vec E\rangle\langle\vec\Sigma,\hat{\vec\CP}\rangle
]_-\biggr\},
$$
\begin{eqnarray*}
[\langle\vec\Sigma,\vec H\rangle,\langle\vec\Sigma,\hat{\vec\CP}\rangle]_-&=&
[\langle\vec H,\hat{\vec\CP}\rangle]_-+ i\langle\vec\Sigma,\vec H\times
\hat{\vec\CP}\rangle- i\langle\vec\Sigma,\hat{\vec\CP}\times\vec H\rangle=\\
&=&2 i\langle\vec\Sigma,\vec H\times\hat{\vec\CP}\rangle+ i\h
\div\vec H-\h\langle\vec\Sigma,\rot\vec H\rangle,
\end{eqnarray*}
as was to be proved.
\begin{Lem} The following relation holds{\rm :}
\begin{eqnarray}
[\hat\CH_1,\hat{\vec S}]&=&{e_0(g-2)\over 2(m_0c)^2}\biggl\{(m_0c)
(-\vec\Sigma\times\vec H-\vec E\rho_1)+\rho_3\langle\vec\Sigma,
\vec E\rangle\hat{\vec\CP}-\rho_2\langle\vec\Sigma,\vec H\rangle\hat{\vec\CP}-{}\cr
&-&{ i\h\over 2}\grad(\rho_3\langle\vec\Sigma,\vec E\rangle+\rho_2\langle\vec\Sigma,
\vec H\rangle)\biggr\}.
\end{eqnarray}
\end{Lem}
{\bf Proof}. Actually,
\begin{eqnarray*}
\lefteqn{[\hat\CH_1,\rho_3\vec\Sigma]_-={ ie_0(g-2)\over 2m_0c}\biggl\{
[\langle\vec\Sigma,\vec H\rangle,\vec\Sigma]_-+ i\rho_1[\langle
\vec\Sigma,\vec E\rangle,\vec\Sigma]_+\biggr\}=}\\
&&={e_0(g-2)\over 2m_0c}\{-\vec\Sigma\times\vec H-\vec E\cdot\rho_1\};\\
\lefteqn{\Big[\hat\CH_1,{1\over m_0c}\rho_1\hat{\vec\CP}\Big]_-={ ie_0(g-2)\over 2m_0c}
\biggl\{ i\rho_2[\langle\vec\Sigma,\vec H\rangle,\hat{\vec\CP}]_+-
 i\rho_3[\langle\vec\Sigma,\vec E\rangle,\hat{\vec\CP}]_+\biggr\}=}\\
&&={e_0(g-2)\over (m_0c)^2}\biggl\{\rho_3\langle\vec\Sigma,\vec E\rangle\hat{\vec\CP}-
\rho_2\langle\vec\Sigma,\vec H\rangle\hat{\vec\CP}-{ i\h\over 2}\grad
(\rho_3\langle\vec\Sigma,\vec E\rangle-\rho_2\langle\vec\Sigma,\vec H\rangle)
\biggr\},
\end{eqnarray*}
as was to be proved.
\begin{Lem} The following relation holds{\rm :}
\begin{equation}
\Pi^t_+(t)\biggl\{{\pa\hat S_0\over\pa t}+{ i\over\h}[\hat\CH_0,\hat S_0]_-
\biggr\}\Pi_+(t)={e\over m_0c\gamma}\biggl\{\langle\vec\sigma,\vec E\rangle+
\gamma\frac{\langle\vec\beta,\vec E\rangle}{\ga}\sb\biggr\}.
\end{equation}
\end{Lem}
{\bf Proof} follows directly from (A.9).
\begin{Lem} The following relation holds{\rm :}
\begin{eqnarray}
\lefteqn{\Pi^t_+(t)\biggl\{{\pa\hat{\vec S}\over\pa t}+{ i\over\h}[\hat\CH_0,\hat{\vec S}]_-
\biggr\}\Pi_+(t)={e\over m_0c}\biggl\{\sb\vec E+{}}\cr
&&{}+\gamma^{-1}\vec H\times\vec\sigma+\frac{1}{\ga}\sb\vec H
\times\vec\beta\biggr\}.
\end{eqnarray}
\end{Lem}
{\bf Proof} follows directly from (A.12) and (A.9).
\begin{Lem} The following relation holds{\rm :}
\begin{eqnarray}
\Pi^t_+(t)[\hat\CH_1,\hat S_0]_-\Pi_+(t)&=&-{e_0(g-2)\over (m_0c)^2}\biggl\{
\biggl(\sb{1\over\ga}\langle\vec\beta,\vec H\times\hat{\vec\CP}\rangle-{}\cr
&-&\langle\vec\sigma,\vec H\times\hat{\vec\CP}\rangle\biggr)-\langle\vec\sigma,
\vec\beta\times(\vec E\times\hat{\vec\CP})\rangle\biggr\}+O(\h).
\end{eqnarray}
\end{Lem}
{\bf Proof} follows directly from (A.7) and (A.8).
\begin{Lem} The following relation holds{\rm :}
\begin{eqnarray}
\lefteqn{\Pi^t_+(t)[\hat\CH_1,\hat{\vec S}]_-\Pi_+(t)={e_0(g-2)\over (m_0c)^2}
\biggl\{-m_0c\biggl(\sb\vec E-\gamma^{-1}\vec H\times\vec\sigma-{}}\cr
&&{}-\frac{1}{\ga}\sb\vec H\times\vec\beta\biggr)-\biggl(\sb\frac{\langle
\vec\beta,\vec E\rangle}{\ga}-\langle\vec\beta,\vec E\rangle-\langle
\vec\sigma,\vec\beta\times\vec H\rangle\biggr)\hat{\vec\CP}\biggr\}+O(\h).
\end{eqnarray}
\end{Lem}
{\bf Proof} follows directly from (B.7), (A.7) and (A.8).
\begin{Lem} The following relations hold{\rm :}
\begin{equation}
\Pi^t_+(t)\hat S_0\Pi_+(t)={1\over m_0c}\biggl(\gamma^{-1}\sp+
\frac{\langle\vec\beta,\hat{\vec\CP}\rangle}{\ga}\sb\biggr),
\end{equation}
\begin{equation}
\Pi_+^t(t)\hat{\vec S}\Pi_+(t)=\vec\sigma-\sb\frac{\vec\beta}{\ga}+
\frac{1}{m_0c}\hat{\vec\CP}\sb.
\end{equation}
\end{Lem}
{\bf Proof} follows directly from (A.9), (A.7) and (A.12).
\begin{Lem} The vector
\[\vec a=\vec\zeta+\frac{\gamma\vec\beta}{\ga}\langle
\vec\zeta,\vec\beta\rangle\]
satisfies the equation
\begin{equation}
\dot{\vec a}={ge\over 2m_0c\gamma}(\langle\vec a,\vec\beta\rangle\vec E+
\vec H\times\vec a)+{e(g-2)\gamma\over 2m_0c}\vec\beta(\langle\vec a,
\vec E\rangle+\langle\vec a,\vec\beta\rangle\langle\vec\beta,\vec E\rangle+
\langle\vec\beta,\vec a\times\vec H\rangle).
\end{equation}
\end{Lem}
{\bf Proof}. We express the vector $\vec\zeta$ in terms of $\vec a$ and obtain
\begin{equation}
\vec\zeta=\vec a-\frac{\vec\beta}{\ga}\langle\vec a,\vec\beta\rangle
\end{equation}
(see (A.1) and (A.2)). Averaging ('.9) with respect to $J^{(1)}$ and setting
$\h\to0$, we substitute ('.15) into the relation obtained. Then
\begin{eqnarray*}
\displaystyle{e\over m_0c}\biggr\{\gamma^{-1}\langle\vec a,\vec\beta\rangle\vec E+
\gamma^{-1}\vec H\times\biggr(\vec a-\frac{\vec\beta}{\ga}\langle\vec a,
\vec\beta\rangle\biggr)&+&\frac{1}{\ga}\gamma^{-1}\langle\vec a,
\vec\beta\rangle\vec H\times\vec\beta\bigg\}=\\
&=&{e\over m_0c\gamma}\{\langle\vec a,\vec\beta\rangle
\vec E+\vec H\times\vec a\}.
\end{eqnarray*}
Similarly, we transform ('.11)
\begin{eqnarray*}
\lefteqn{{e(g-2)\over (m_0c)^2}\biggr\{-m_0c(\langle\vec a,
\vec\beta\rangle\vec E+\vec H\times\vec a)\gamma^{-1}-{}}\\
&&{}-\gamma m_0c\vec\beta\biggl(\gamma^{-1}\langle\vec a,\vec\beta\rangle
\frac{\langle\vec\beta,\vec E\rangle}{\ga}+\langle\vec a,\vec E\rangle+
\frac{\langle\vec\beta,\vec E\rangle}{\ga}\langle\vec a,\vec\beta\rangle-
\langle\vec a,\vec\beta\times\vec H\rangle\biggl)={}\\
&&{}={e(g-2)\over 2m_0c}\{\gamma^{-1}(\langle\vec a,\vec\beta\rangle\vec E+
\vec H\times\vec a)+\gamma\vec\beta(\langle\vec a,\vec E\rangle+
\langle\vec a,\vec\beta\rangle\langle\vec\beta,\vec E\rangle-
\langle\vec a,\vec\beta\times\vec H\rangle)\}.
\end{eqnarray*}
Substituting the latter relation into (2.30), we get ('.14), as was to be
proved.

\end{document}